\documentclass[10pt,journal,compsoc]{IEEEtran}
    
	\usepackage{amssymb}
	\usepackage{graphicx}
	\usepackage{epstopdf}
	\usepackage{gensymb}
	\usepackage{color}
	\usepackage{amsmath}
	\usepackage{enumerate}
	\usepackage{cite}
	\usepackage{comment}
	\usepackage{amsthm}
	\usepackage{enumitem}
	\theoremstyle{definition}
	\usepackage{soul}
	
	\usepackage[framemethod=TikZ]{mdframed}
	
	\usepackage{algorithmicx}
	\usepackage[ruled]{algorithm}
	\usepackage{algpseudocode}
	\alglanguage{pseudocode}
    
	\usepackage{array}
	\usepackage{booktabs}
	\usepackage{multirow}

	\usepackage{threeparttable} 
	\usepackage{hyperref}
	
	\usepackage{mathtools}

	\usepackage{url}
	\urldef{\mailsa}\path|{alfred.hofmann, ursula.barth, ingrid.haas, frank.holzwarth,|
		\urldef{\mailsb}\path|anna.kramer, leonie.kunz, christine.reiss, nicole.sator,|
		\urldef{\mailsc}\path|erika.siebert-cole, peter.strasser, lncs}@springer.com|

\definecolor{garrisonpink1}{rgb}{0.858, 0.188, 0.478}

\newcommand{\mypara}[1]{\vspace{2pt}\noindent\textbf{{#1: }}}
\definecolor{PangLHpink2}{rgb}{0.5, 0.5, 1}

\definecolor{yjr}{rgb}{0.977, 0, 0}

\begin{document}

\title{Design and Evaluate Recomposited \underline{O}R-\underline{A}ND-\underline{X}OR\underline{-PUF}}

\author{
Jianrong Yao, Lihui Pang, Yang Su, Zhi Zhang, Wei Yang, Anmin Fu, and Yansong Gao.

\IEEEcompsocitemizethanks{\IEEEcompsocthanksitem J. Yao, Y. Gao, W. Yang, and A. Fu are with School of Computer Science and Engineering, Nanjing University of Science and Technology, Nanjing, China. e-mail: \{120106222744,yansong.gao;generalyzy;fuam\}@njust.edu.cn.}

\IEEEcompsocitemizethanks{\IEEEcompsocthanksitem L. Pang is with School of Electrical Engineering, University of South China, Hengyang, China, and security engineering laboratory, Sungkyunkwan University, Korea. e-mail: sunshine.plh@hotmail.com}

\IEEEcompsocitemizethanks{\IEEEcompsocthanksitem  Y. Su is with the School of Computer Science, The University of Adelaide, Adelaide, Australia. e-mail: yang.su01@adelaide.edu.au}

\IEEEcompsocitemizethanks{\IEEEcompsocthanksitem Z. Zhang is with Data61, CSIRO, Sydney, Australia. e-mail: zhi.zhang@data61.csiro.au.}

\IEEEcompsocitemizethanks{\IEEEcompsocthanksitem Corresponding Author: Y. Gao.}

}

\IEEEtitleabstractindextext{		
\begin{abstract}
Physical Unclonable Function (PUF) is a hardware security primitive with a desirable feature of low cost. Considering the space of challenge-response pairs (CRPs), there are two PUF categories: weak PUF and strong PUF. Compared to a weak PUF, a strong PUF has a wider range of applications. However, it is challenging to design a \textit{reliable and secure lightweight} strong PUF. To address this challenge, PUF recomposition built upon multiple simple PUF instances has received a lot of attention in research, such as the most popular XOR-APUF, the recent MPUF in IEEE TC 2017~\cite{sahoo2017multiplexer}, XOR-FF-APUF in IEEE TIFS 2020~\cite{avvaru2020homogeneous} and IPUF in TCHES 2020~\cite{nguyen2019interpose}.

When a combination of MAX and MIN (equal to AND and OR) bitwise operations are used in PUF recomposition~\cite{ruhrmair2010modeling,ruhrmair2013puf,gao2020physical}, its resilience against model attacks was expected to be improved markedly, because one bitwise operation might be vulnerable to one type of modeling attack and combining them can yield improved resilience. 
To our knowledge, there is no explicit evaluation of this recomposition; thus, this study is the first to evaluate the 
uniformity and reliability of the \underline{O}R-\underline{A}ND-\underline{X}OR\underline{-PUF} (OAX-PUF)---($x,y,z$)-OAX-PUF. Compared to the most used $l$-XOR-PUF, the ($x,y,z$)-OAX-PUF  shows better reliability given $l=x+y+z$ without degrading the uniformity (i.e., retain to be 50\%). As APUF is a compact PUF instance for constructing lightweight strong PUF candidates, e.g., XOR-APUF, MUXPUF and IPUF, we further examine the modeling resilience of the ($x,y,z$)-OAX-APUF using four powerful attacks, i.e., logistic regression (LR), reliability assisted CMA-ES, multilayer perceptron (MLP), and the most recent hybrid LR-reliability.
Compared to the XOR-APUF, the OAX-APUF successfully defeats the CMA-ES attack. It often shows improved modeling resilience against LR and hybrid LR-reliability attacks while always increasing the attacking time costs of these two attacks. However, OAX-APUF exhibits lower modeling resilience against the MLP attack unless the $x,y,z$ are carefully tuned. Overall, the OAX recomposition could be an alternative lightweight recomposition approach in constructing strong PUFs if the underlying PUF( e.g., FF-APUF), has shown improved resilience against modeling attacks, because the OAX incurs smaller reliability degradation compared to XOR.
\end{abstract}

\begin{IEEEkeywords}
Recomposited PUF, XOR-PUF, OR-AND-XOR-PUF, Modeling Attacks.
\end{IEEEkeywords}}

\maketitle

\section{Introduction}\label{Sec:Intro}

A physical unclonable function (PUF) leverages inevitable randomness induced from hardware manufacturing processes to give each hardware instance a unique ``fingerprint"~\cite{gassend2002silicon,herder2014physical,gao2020physical}, making itself a hardware security primitive. 
Because over 20 billion Internet of Things (IoT) devices were deployed before 2021, and this number is still increasing~\cite{GarnerPrediction}, it is critical to use a lightweight security mechanism in the pervasively deployed IoT devices~\cite{chang2017retrospective,gao2020physical}. Thus, PUFs (silicon PUFs in particular),  which are characterized by their low cost, have received a lot of attention from both academia and industry.

As a function, PUF is treated as a black-box with its inputs and outputs, and we can characterize it by CRPs. Particularly, a PUF-instance-dependent response (output) is stimulated due to a challenge (input). According to the CRP space, PUF can be categorized into two classes: weak PUF and strong PUF. A weak PUF has a limited number of CRPs and is primarily used for provisioning cryptographic keys. A well-known weak PUF is the memory-based PUF (e.g., SRAM PUF uses an SRAM cell's address as a challenge and treats the power-up pattern of the cell as the response~\cite{gao2018lightweight}). The challenge-response interface should be protected for weak PUFs to prevent exhaustive CRP characterization.
Compared to a weak PUF, a strong PUF has a wider range of applications and does not require  challenge-response interface protection. For example, a strong PUF enables lightweight authentication by employing one CRP a time, similar to a one-time pad~\cite{gao2017puf}. Additionally, a strong PUF is applicable for virtual proofs of reality~\cite{ruhrmair2015virtual} and advanced cryptographic protocols~\cite{brzuska2011physically,ruhrmair2012practical,ruhrmair2013practical,ruhrmair2013pufs}, such as oblivious transfer, bit commitment and multiparty computation.

However, designing a reliable and secure lightweight strong PUF is challenging and is actually a fundamental open problem of strong silicon PUFs~\cite{gao2020physical}. Specifically, a strong PUF instance faces security threats from modeling attacks where an attacker uses collected CRPs and models the strong PUF using machine learning (ML) techniques to accurately predict a response given an unseen challenge. To improve resilience to modeling attacks, PUF recomposition has been proposed by organizing a number of basic PUF instances, which is similar to the case of secure cipher construction where diffusion and confusion are used in many rounds, given that one single round is insecure. 

A representative single strong PUF is the Arbiter PUF (APUF), which can generate a larger CRP space, compared to the weak PUF. However, an APUF instance can be easily broken by modeling attacks. To improve resilience, there have been a number of proposed APUF recompositions, such as XOR-APUF, MUXPUF~\cite{sahoo2017multiplexer} and IPUF~\cite{nguyen2019interpose}. These PUF recompsitions do not apply additional security blocks (e.g., random number generator (RNG), hash function, etc.~\cite{gassend2008controlled,gao2016obfuscated,yu2016lockdown,herder2016trapdoor,gao2017puf,9091218} and leave an open challenge-response interface, which makes a recomposited strong PUF can be equally accessed by any party and serve as a transparent hardware security primitive. Thus, a recomposited PUF can still be used in various applications without limiting itself to only lightweight authentication and key generation~\cite{gao2016obfuscated,yu2016lockdown,herder2016trapdoor,gao2017puf,9091218}.
Essentially, PUF recomposition injects nonlinearity into the overall PUF structure to mitigate the modeling attacks. However, one primary issue of nonlinearity injection is reliability degradation for a recomposited PUF, and it is nontrivial to recompose many basic PUF instances while retaining  applicable reliability. Thus, it is important to retain the reliability as much as possible during recomposition.

\mypara{OAX-PUF} Leveraging MAX and MIN (equal to AND and OR) bitwise operations is  likely a promising lightweight recomposition approach~\cite{ruhrmair2010modeling,ruhrmair2013puf,gao2020physical}, because one bitwise operation-based PUF recomposition might be vulnerable to one specific type of modeling attack and combining them can have an improved overall modeling resilience. To date, there is no concrete design and evaluation of PUF recomposition taking the AND and OR into consideration.

While the AND and OR bitwise operations are \textit{simple and lightweight} in recompositing strong PUFs, their \textit{advantages} and \textit{limitations} are still unclear to the PUF community. In this work, we propose \underline{O}R-\underline{A}ND-\underline{X}OR\underline{-PUF} (OAX-PUF)---($x,y,z$)-OAX-PUF by incorporating three basic logic operations, where $x,y,z$ are the number of PUF instances for OR, AND, and XOR bitwise operations, respectively. To comprehensively evaluate the ($x,y,z$)-OAX-PUF built upon the operations, we aim to answer the following two research questions (RQs):

\vspace{-2mm}
\begin{mdframed}[backgroundcolor=black!10,rightline=false,leftline=false,topline=false,bottomline=false,roundcorner=2mm]
	\textbf{RQ1:} What are the reliability and uniformity of the ($x,y,z$)-OAX-PUF? 
\end{mdframed}
\vspace{-3mm}
\begin{mdframed}[backgroundcolor=black!10,rightline=false,leftline=false,topline=false,bottomline=false,roundcorner=2mm]
	\textbf{RQ2:}  What is the modeling resilience of the ($x,y,z$)-OAX-PUF?
\end{mdframed}
\vspace{-3mm}

To answer \textbf{RQ1}, we formulate the reliability and uniformity of the ($x,y,z$)-OAX-PUF and the formulations are \textit{independent} of the underlying PUF instances that are used. To answer \textbf{RQ2}, we take the \textit{APUF as an underlying PUF} to build up the ($x,y,z$)-OAX-APUF considering that the APUF is a mainstream basic PUF to construct recomposited PUFs. We then comprehensively examine the ($x,y,z$)-OAX-APUF  resilience against four powerful modeling attacks, including logistic regression (LR), reliability assisted CMA-ES, MLP, and the most recent hybrid LR-reliability. 
Considering that XOR-APUF is the most studied APUF variant, we compare it with the proposed OAX-APUF in modeling resilience.

\mypara{Our Contributions} The primary contributions and results  of this work are summarized as follows.
\begin{itemize}[noitemsep, topsep=2pt, partopsep=0pt,leftmargin=0.4cm]
  \item To our knowledge, we are the first to explore the applicability of using all three basic logic operations to add nonlinearity into recomposited PUFs,
    where prior works focused on the bitwise logic operation of XOR.

    \item We formulate critical performance metrics for the OAX-PUF (i.e., reliability and uniformity) and perform an extensive evaluation of the OAX-PUF to validate the efficacy of the formulations. These formulations are \textit{independent} of underlying PUFs and thus allow further improvements on the OAX-PUF and its variants. \textbf{RQ1}.
    
    \item We quantitatively evaluate the ($x,y,z$)-OAX-PUF and compare it with the $l$-XOR-PUF in terms of the formulated metrics, the results of which show that both have uniformity near an ideal value of 50\%. The OAX-PUF also exhibits enhanced reliability when both use the same number of underlying PUFs (i.e., $x+y+z=l$ ). \textbf{RQ1}.
    
    \item We use \textit{APUF as an underlying PUF instance} and further examine OAX-APUF's resilience against four powerful modeling attacks, the results of which are compared with those of XOR-APUF. In particular, OAX-APUF has the advantage of resisting to the reliability assisted modeling attacks, including CMA-ES and hybrid LR-reliability attacks~\cite{tobisch2021combining}---it even successfully defeats the former CMA-ES attack. The OAX-APUF also increases attacking time costs against LR attack when the modeling resilience is normally no less than XOR-APUF particularly on a large-scale. However, the OAX-APUF does not show improved resilience against the MLP attack. \textbf{RQ2}.

\end{itemize}

\mypara{Paper Organization} 
In Section~\ref{sec:background}, we provide background on APUF, XOR-APUF and four types of modeling attacks. In Section~\ref{sec:oax-apuf}, we present the design of OAX-PUF, formulate two important performance metrics and validate the formulations by numerical experiments and silicon measurements. In Section~\ref{Sec:securityEvaluations}, using APUF as a basic PUF for a case study, we examine and compare OAX-APUF with XOR-APUF in terms of modeling resilience against four powerful modeling attacks.
Section~\ref{sec:relatedwork} presents related works of recomposited PUFs. We also discuss the OAX-PUF and some means of improving its modeling attack resilience in Section~\ref{sec:discussion} by incorporating observations from the current work.
We conclude this work in Section~\ref{sec:conclusion}.

\section{Background}\label{sec:background}
In this section, we first introduce the Arbiter-based PUF that we will use as a case study to build the OAX-PUF and one of its most popular variants, XOR-APUF, to which we will compare the OAX-PUF. We then present the four most powerful modeling attacks on XOR-APUF. The former two attacks solely use the CRPs, while the third attack uses one type of side channel information that is the unreliability to assist the APUF model building. The last attack uses both response and reliability information. It is possible to use other side channel information such as power, timing~\cite{ruhrmair2014efficient} and photonic~\cite{tajik2017photonic} to facilitate APUF modeling attacks, but they are expensive to collect compared with the reliability information. For example, timing and photonic information require accurate equipment or good expertise. In addition, physical access to the PUF device is also typically required. We primarily focus on CRP based and unreliability (or challenge reliability pair) based modeling attacks, which are cheaper and most attractive to attackers in the real-world. This process is common and aligned with other recent studies~\cite{nguyen2019interpose,zhang2020set,avvaru2020homogeneous} when evaluating PUF modeling resilience.

\subsection{Arbiter-based PUF}
The APUF exploits manufacturing variabilities that result in random interconnect and transistor gate time delays~\cite{gassend2002silicon}. This structure is simple, compact, and capable of yielding a large CRP space. Unlike the optical PUF~\cite{pappu2002physical}, the APUF is, however, built upon a linear additive structure and thereof vulnerable to various modeling attacks where an attacker uses known CRP observations to build a model that predicts responses accurately given yet seen challenges~\cite{ruhrmair2010modeling,ruhrmair2013puf,becker2015gap,becker2015pitfalls}. 

\mypara{$\textbf{Linear Additive Delay Model}$}
A linear additive delay model of APUFs can be expressed as~\cite{lim2005extracting}:
  \begin{equation}
      \Delta={\vec{w}^{T}\vec{\Phi}} \label{con:delta},
  \end{equation}
where $\vec{w}$ is the weight vector that models the time delay segments in the APUF, and $\vec{\Phi}$ is the parity (or
feature) vector that can be generally understood as a transformation of the challenge. The dimension of both $\vec{w}$  and $\vec{\Phi}$ is $n+1$ given an $n$-stage APUF.
 \begin{equation}
     \Phi[n]=1,\Phi[i]==\prod_{j=i}^{n-1}{(1-2c[j])},i=0,...,n-1. \label{con:feature vector}
  \end{equation}

The response of an $n$-stage APUF is determined by the delay difference between the top and bottom signals. This delay difference is the sum of the delay differences of each of the individual $n$ stages. The delay difference of each stage depends on the corresponding challenge\cite{becker2015gap}. Based on Eq.~\ref{con:delta}, the response $r$ of the challenge \textbf{c} is modelled as:
 \begin{equation}
    r=\begin{cases}
     1, \text{ if } \Delta<0 \\ 
     0, \text{otherwise} .
 \end{cases}
 \label{con:responses}
 \end{equation}

\mypara{$\textbf{Short-Time Reliability}$}
Due to noise, environmental parameter changes, and aging, the reproducibility or reliability of the PUF response is not perfect (i.e., feeding the same challenge to a PUF cannot always produce the fully stable response). 

From the security perspective, the reliability of APUFs leaks information that can be used to infer the APUF internal time delay. For example, if a given response is unreliable, then $\Delta$ in Eq.~\ref{con:delta} given this response is near $0$. Such unreliability is a type of side-channel information that can be easily captured by an attacker to build the APUF model~\cite{delvaux2013side,becker2015gap}. The short-time reliability R (i.e., repeatability) merely resulting from the noise for a specific challenge $\bf c$ of the APUF can be easily measured by the following means: we assume that the challenge $\bf c$ is applied $M$ times, and that there are $N$ '1' response value occurrences out of $M$ repeated measurements. The short-time reliability is expressed as~\cite{nguyen2019interpose}:
\begin{equation}\label{eq:shortReliability}
    R = N/M \in[0, 1].
\end{equation}

As easy-to-obtain side-channel information, R, can be exploited by an attacker to perform reliability-based modeling attacks.

\begin{figure}
	\centering
	\includegraphics[trim=0 0 0 0,clip,width=0.4\textwidth]{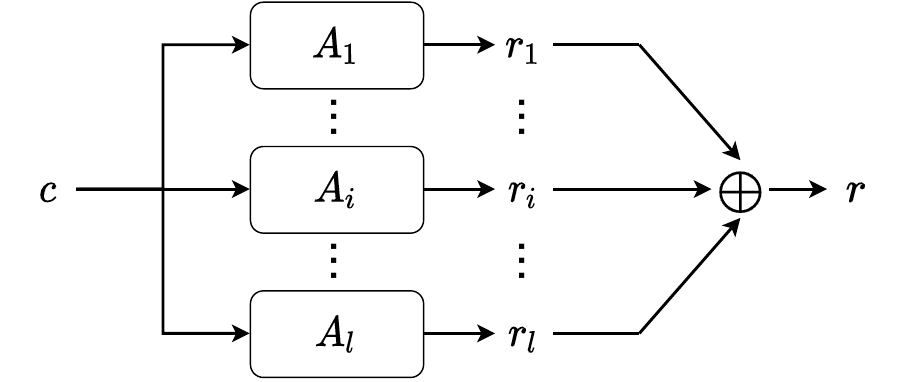}
	\caption{$l$-XOR-APUF consists of $l$ APUFs and each of the APUF response, $\{r_1,...,r_l\}$ is XORed at the end to form a 1-bit response $r$. All APUFs share the same challenge \textbf{c}.}
	\label{fig:xor}
\end{figure}
\subsection{XOR-APUF}
As shown in Fig.\ref{fig:xor}, $l$-XOR-APUF is constructed by recompositing $l$ APUFs in parallel from the topology perspective. Each APUF shares the same challenge and produces a digital response. All $l$ responses are XORed to form the final $l$-XOR-APUF response. Using a larger $l$ can nearly exponentially increase the modeling attack complexity when \textit{only CRP is used}---in this context, the most efficient ML attack is LR according to~\cite{nguyen2019interpose}. It is suggested that a $10$-XOR-APUF is secure against LR attack when the APUF is with 64 stages~\cite{tobisch2015scaling}. However, the $l$-XOR-APUF becomes severely unreliable when $l$ is large (e.g., 10), which negatively restricts the large $l$ usage. In addition, the large $l$ of a $l$-XOR-APUF is still ineffective against reliability-based modeling attacks---it uses reliability information of the CRP---since the complexity of such attack is only linearly increased as a function of $l$.

\subsection{Modeling Attacks}\label{sec:modelattack}
There are four prominent modeling attacks against the APUF and its variants, that is, logistic regression (LR), multilayer perceptron (MLP), covariance matrix adaptation evolution strategy (CMA-ES) and hybrid LR-reliability attack.

\mypara{Logistic Regression}
 LR is a common classification and prediction algorithm, that is primarily used for classification problems.  LR describes the relationship between the independent variable $x$ and the dependent variable $y$, and then predicts the dependent variable $y$. We assume that the hypothetical function is:
\begin{equation}
h_{\theta }(x)=g(\theta ^{T}x),~
g(f)=\frac{1}{1+e^{f}},
\end{equation}
where $g$ is a sigmoid function used by the LR for classification, $x$ is the input, and $\theta$ is the parameter to be solved in logistic regression. Then the decision boundary is:
\begin{equation}
f=\theta ^{T}x=0.
\end{equation}

The LR first fits the decision boundary and then establishes the probabilistic relationship between the boundary and classification. Binary classification assumes that the data follow a Bernoulli distribution. Using the maximum likelihood function, gradient descent is used to solve the parameter $\theta$ to obtain the classification probability, and the sigmoid function is used for ultimate binary class decision. If $\theta ^{T}x \geq 0$, $y$=1; otherwise, $y$= 0.
The LR is used to break APUFs and XOR-APUFs early in 2010~\cite{ruhrmair2010modeling}. Because the response of a PUF is `0' or `1', modeling attack on PUFs via LR is thus a binary classification problem. For different PUFs, the decision boundary can vary. According to~\cite{ruhrmair2010modeling}, the decision boundary of the APUF is expressed as:

\begin{equation}
f={\vec{w}^{T}\vec{\Phi}}=0.
\end{equation}

The decision boundary of $l$-XOR-APUF is expressed as:

\begin{equation}
f=\prod_{i=1}^{l}{\vec{w}^{T}_{i}\vec{\Phi _{i}}}=0,
\end{equation}
where $\vec{w_{i}}$ and
$\vec{\Phi_{i}}$ denote the parameter and feature vector, respectively, for
the $i$-th APUF.

Rührmair \textit{et al.} \cite{ruhrmair2010modeling} showed that the RPROP gradient descent algorithm is the most effective method for LR attacks on PUFs. Johannes Tobisch and Georg T. Becker~\cite{tobisch2015scaling} showed later that when the stage of $l$-XOR-APUF is 64, $l$ is within 9, the $l$-XOR-APUF can be successfully modeled with high accuracy. When the stage is 128, $l$ is within 7, the $l$-XOR-APUF can also be broken.

\begin{figure}
	\centering
	\includegraphics[trim=0 0 0 0,clip,width=0.4\textwidth]{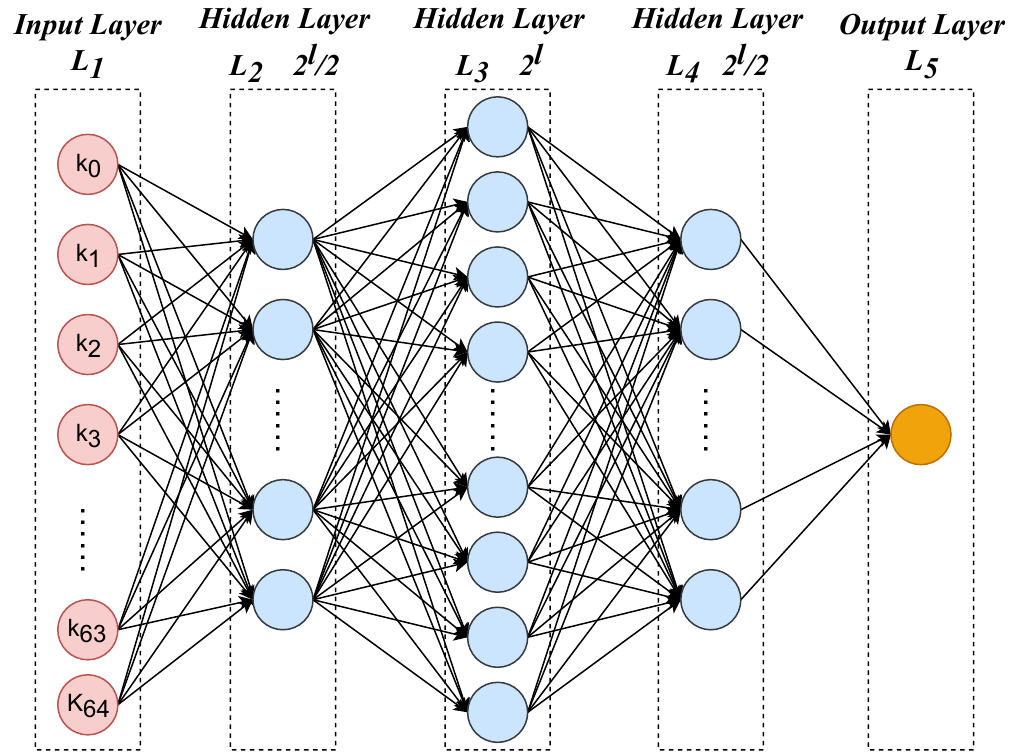}
	\caption{  MLP architecture proposed by Mursi \textit{et al.}~\cite{mursi2020fast}. The MLP architecture has five layers, among which there are three hidden layers, and the activation function of the hidden layers is \textsf{tanh}. The input of the input layer is the feature vector corresponding to the challenge, and the activation function of the output layer is \textsf{sigmoid}.
 }
	\label{fig:MLP}
\end{figure}

\mypara{Multilayer Perceptron} MLP has been used to model strong PUFs since 2012. Hospodar \textit{et al.}~\cite{hospodar2012machine} used an MLP architecture with a hidden layer of four neurons to model a 2-XOR-APUF with 64 stages, with a prediction accuracy of approximately 90\%. In 2017, Alkatheiri \textit{et al.}~\cite{alkatheiri2017towards} proposed a new MLP architecture with three hidden layers, and each hidden layer has $2^{k+1}$ neurons, where $k$ is the number of loops in an FF-PUF. This MLP architecture can break FF-APUF with 64 and 128 stages. Then, several MLP architectures were used to model $l$-XOR-APUFs. Among them, the MLP architecture of Mursi \textit{et al.}~\cite{mursi2020fast} is the most effective one. Similar to LR, the input of the MLP is the feature vector corresponding to the challenge. As shown in  Fig.~\ref{fig:MLP}, the MLP architecture has three hidden layers. The number of neurons in the first hidden layer and the third hidden layer is $2^{l}/2$, while the number of neurons in the second hidden layer is $2^{l}$, where $l$ means the number of APUFs. In terms of the attack effect, compared with the previous MLP, this MLP architecture can successfully model $l$-XOR-APUF with less time and CRPs. Considering that \textsf{tanh} can be accurately approximate the binary function of APUF and that using \textsf{tanh} can reduce the number of hidden layers, they choose it as the activation function of the hidden layers.

\mypara{CMA-ES}\label{sec:ESattack}
Classical ML attacks against APUFs rely solely on the challenge response pairs, which distinguishes them from reliability assisted CMA-ES attacks. CMA-ES attacks exploit a response's reliability of a targeted APUF. The response's reliability refers to how often the APUF generates the same response for a given challenge~\cite{becker2015gap}. The CMA-ES attack against APUFs has the following six steps~\cite{tobisch2015scaling}:
\begin{itemize}[noitemsep, topsep=2pt, partopsep=0pt,leftmargin=0.4cm]
\item
Collect N challenges  
$C=\{\boldsymbol{c}_{1},...,\boldsymbol{c}_{i},...,\boldsymbol{c}_{N}\}$, and use Eq.~\ref{con:feature vector} to gain $c_{i}$'s corresponding feature vector $\vec{\Phi_{i}},(i=1,2,...,N)$.
Thus, many CRPs can be collected to test whether the CMA-ES modeling converges.
\item
 Send the same $\vec{\Phi_{i}}$ to the APUF $m$ times and collect $m$ repeated responses: $r_{i,1},r_{i,2},...r_{i,m}$. The  reliability $h_{i}$ of each challenge $\boldsymbol{c}_{i}$ is calculated and generates the reliability vector $h = (h_{1},..., h_{i},..., h_{N})$. The reliability is calculated as follows \cite{becker2015gap}:
    \begin{equation}
        h_{i}=|\frac{m}{2}-\sum_{j=1}^{m}r_{i,j}|.
    \end{equation}
\item 

Generate $K$ APUF models:\\
$\{(\vec{w_{1}},\epsilon_{1}),...,(\vec{w_{j}},\epsilon_{j}),...,(\vec{w_{K}},\epsilon_{K})\}$~\cite{nguyen2019interpose}. For each APUF model, there are two steps as follows:
\begin{itemize}[noitemsep, topsep=2pt, partopsep=0pt,leftmargin=0.4cm]
    \item 
    The hypothetical reliability $\widetilde{h_{i}}$ corresponding to each challenge $\boldsymbol{c}_{i}$ is calculated to form the hypothetical reliability vector $\widetilde{h}=(\widetilde{h_{1}},...,\widetilde{h_{i}},...,\widetilde{h_{N}})$. The calculation formula of hypothetical reliability is expressed as~\cite{becker2015gap}, where $\epsilon$ is an error boundary that must be approximated by the machine learning algorithm:
    \begin{equation}
        \widetilde{h_{i}}=\begin{cases}
       1, \text{ if } |\vec{w}^{T}\vec{\Phi_{i}}|\ge \epsilon \\ 
       0, \text{ if } |\vec{w}^{T}\vec{\Phi_{i}}|< \epsilon  ,
    \end{cases}
    \label{eq.9}
    \end{equation}
where $\epsilon$=$\epsilon_{j},\vec{w}=\vec{w_{j}},(j=1,2,...,K)$.
    \item Calculate the Pearson correlation between  $\widetilde{h}$ and $h$.
\end{itemize}

\item 

Compare the Pearson correlation of each APUF model and choose $L$ APUF models with the highest Pearson correlation as the parent instances of the next evolution. From these $L$ models, another $K$ child models are generated based on the CMA-ES algorithm. 
\item 
Repeat steps 3 and 4 for $T$ iterations. A model that has the highest Pearson correlation will be selected as the desired model.
\item 

Test CRPs are used to test the model.
If the test accuracy is more than 90\% or less than 10\% (the latter occurs when the model treats `1'/`0' as `0'/`1'), the CMA-ES modeling attack against the APUF succeeds.
\end{itemize}

The main idea behind the reliability assisted CMA-ES attack on $l$-XOR-APUF is that each underlying APUF has equal contributions to the reliability of $l$-XOR-APUF. Becker \textit{et al.}~\cite{becker2015gap} then used a divide-and-conquer strategy to attack $l$-XOR-APUF with CMA-ES based on reliability. There are two cases: the first one is that the same challenge set is used for all APUFs. In this context, the CMA-ES attack can be repeatedly run $l$ times and theoretically converge to a different APUF in the $l$-XOR-APUF each time with the same probability. The second  exploits a different challenge set for each APUF, and then attackers can model a specific APUF.

\mypara{Hybrid LR-reliability Attack} The main idea behind the hybrid or combination of response and reliability attacks is to prevent reliability based attacks from repeatedly converging to the same APUF. As reported by Tobisch \textit{et al.}~\cite{tobisch2021combining}, there are two main ways to introduce weight constraints into reliability based attacks: the first is to retain the original CMA-ES attack and add a new APUF to the weight set in each iteration, which is different from other APUFs in the set; the second is to learn all APUF weights econcurrently and keep them different. The hybrid LR-reliability attack used the second method. The loss function it uses is as follows~\cite{tobisch2021combining}:
\begin{equation}
    \begin{aligned}
    {\rm model}_{\boldsymbol{w}}^{\rm Arbiter}(\boldsymbol{\phi})=|\boldsymbol{w\phi}|,
    \end{aligned}
\end{equation}
\begin{equation}
    \begin{aligned}
    {\rm loss}_{\rm PC}(\boldsymbol{a},\boldsymbol{b})=\frac{\textsf{cov}(\boldsymbol{a},\boldsymbol{b})}{\sqrt{\textsf{ var}(\boldsymbol{a})\cdot \textsf{var}(\boldsymbol{b})}},
    \end{aligned}
\end{equation}
\begin{equation}
\begin{aligned}
    &\text{loss}_{\rm combined}^{XOR}=\varepsilon _{1}^{xor}\sum_{j}\textsf{loss}_{\rm bin}(\textsf{model}_{\boldsymbol{W}}^{\rm total}(\boldsymbol{\Phi} [j,:]),\boldsymbol{r}[j])\\
               &-\varepsilon _{2}^{xor}\sum _{j_{1}}\sum _{j_{2}}\textsf{loss}_{\rm PC}(\textsf{model}_{\boldsymbol{W}[j_{1},:]}^{\rm Arbiter}(\boldsymbol{\Phi} [j_{2},:]),\boldsymbol{h}[j_{2}])\\
               &+\varepsilon _{3}^{xor}\sum_{j_{1}=1}^{k-1} \sum _{j_{2}=j_{1}+1}^{k}|\textsf{loss}_{\rm PC}(\boldsymbol{W}[j_{1},:],\boldsymbol{W}[j_{2},:])|,
\end{aligned}
\label{eq:Hybrid LR-reliability loss}
\end{equation}
where the $\boldsymbol{\Phi}$ represents feature vector matrix, \textbf{r} is the response vector, and \textbf{h} is the corresponding reliability vector. The ${\rm model}^{\rm Arbiter}$ returns the reliability measure of a single APUF, while ${\rm model}^{\rm total}$ describes the complete PUF and outputs a response probability. The $\textsf{loss}_{\rm PC}$ is a loss function based on Pearson correlation. The high prediction accuracy of the model is guaranteed by the first term of Eq.\ref{eq:Hybrid LR-reliability loss}. The second term of Eq.\ref{eq:Hybrid LR-reliability loss} encourages a high correlation between a single APUF and the final output response, while the third term of Eq.\ref{eq:Hybrid LR-reliability loss} limits the similarity between APUFs. $\epsilon_{1}^{xor}$,$\epsilon_{2}^{xor}$ and $\epsilon_{3}^{xor}$ need to be adjusted according to different attack scenarios. A good choice for the value of the constants depends both on PUF parameters, training
set size, noise level and the numerical implementation of the attack itself ~\cite{tobisch2021combining}.

Overall, the hybrid LR-reliability attack is similar to LR except for the loss function devised and used. The loss function avoids the convergence to the same APUF instance, e.g., might be with least unreliability contribution to the recomposited XOR-APUF, compared to the CMA-ES attack, which both leverage reliability information as a side-channel.

\section{OAX-PUF}\label{sec:oax-apuf}
We first present OAX-PUF. Then, two critical metrics consisting of reliability and uniformity are formalized and evaluated. These two formulated metrics are generic and independent of the underlying specific PUF types for constructing the OAX-PUF. 

\begin{figure}
	\centering
	\includegraphics[trim=0 0 0 0,clip,width=0.5\textwidth]{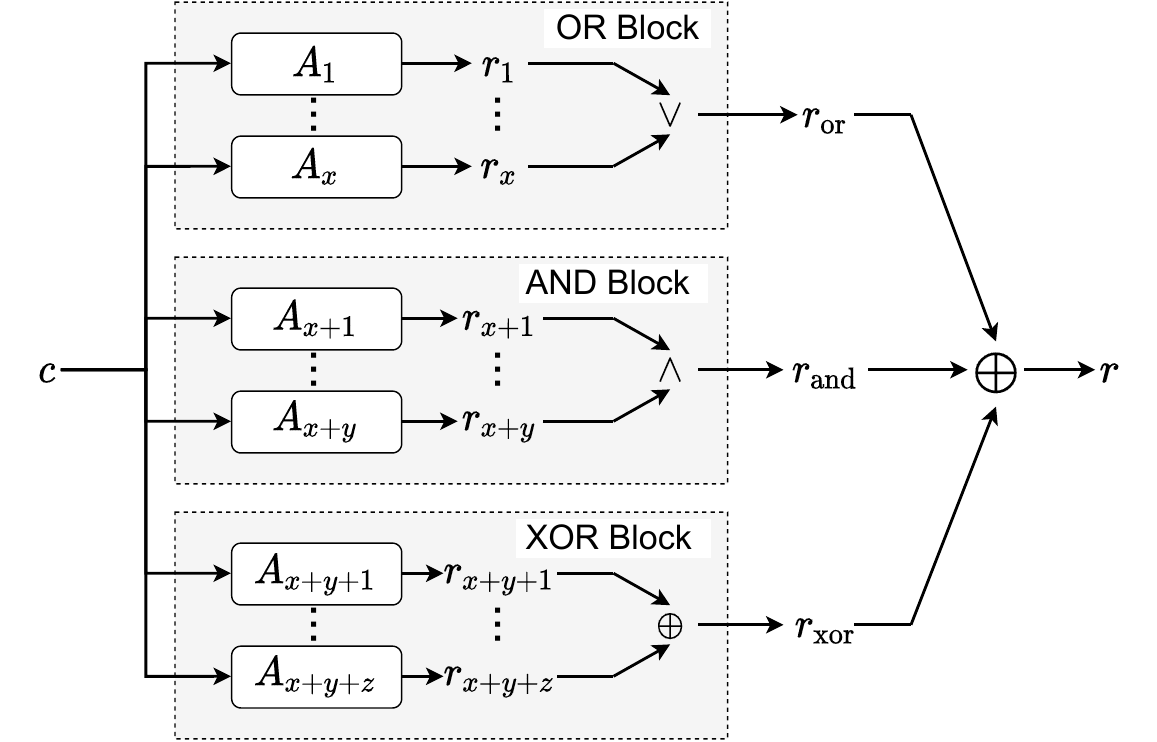}
	\caption{Overview of ($x,y,z$)-OAX-PUF, which has three blocks: OR, AND, and XOR blocks. 
	}
	\label{fig:confidence}
\end{figure}

\subsection{Overview}

Generally, the presented ($x,y,z$)-OAX-PUF is composed of three PUF blocks, as shown in Fig.~\ref{fig:confidence}. The first is the OR-PUF block, in which the responses of $x$ PUFs are ORed to form a 1-bit response $r_{\rm or}$. The second is the AND-PUF block, in which the responses of $y$ PUFs are ANDed, forming a 1-bit response $r_{\rm and}$. The third is the XOR-PUF block, in which the responses of $z$ PUFs are XORed, forming a 1-bit response $r_{\rm xor}$. At the end of ($x,y,z$)-OAX-PUF, $r_{\rm or}$, $r_{\rm and}$ and $r_{\rm xor}$ are XORed to produce the ultimate response $r$ of ($x,y,z$)-OAX-PUF. 

We first formulate the reliability and uniformity of the OAX-PUF. The reliability is related to the usability of a PUF in practice. A PUF should be stable or have high reliability to ease its applications such as lightweight authentication. Uniformity is related to the uniqueness of the PUF. A good uniformity (e.g.,50\%), can easily distinguish one PUF instance from a large population, which is important in, e.g., identification and authentication.

\subsection{Reliability}
PUF unreliability, where $\rm reliability = 1-unreliability$, can be measured using averaged intra-chip Hamming distance (HD) among $m$ repeatedly measured samples of PUF response vectors of length $n$~\cite{Maiti11asystematic}. The intra-chip HD, which is also referred to as as the bit error rate (BER), is expressed as~\cite{Maiti11asystematic}:
\begin{equation}
\label{eq:r}
     \text{BER} = \text{HD}_{\rm INTRA}=\frac{1}{m}\sum_{t=1}^{m}\frac{\text{HD}(\textbf{Res},\textbf{Res}_{t})}{n}\times 100\%,
\end{equation}
where HD$_{\rm INTRA}$ is an averaged number of
different bits between a reference response \textbf{Res} and a regenerated response \textbf{Res}$_t$ at time $t$ corresponding to the same challenge. Thus, PUF reliability is (1-BER), and should be close to 100\% in practice.

To simplify the notation in the following formulations, we denote the BER of a PUF as $\beta$. Each PUF within ($x,y,z$)-OAX-PUF shares the same error rate of $\beta$. The BER of $x$-OR-PUF is termed  $\beta_{\rm or}$. Similarly, $\beta_{\rm and}$ and $\beta_{\rm xor}$ denote the BERs of $y$-AND-PUF and $z$-XOR-PUF, respectively.
The ultimate BER of ($x,y,z$)-OAX-PUF is termed $\beta_{\rm oax}$. It is marked that the reliability of ($x,y,z$)-OAX-PUF is simply  $1-\beta_{\rm oax}$.

We first derive $\beta_{\rm or}$, $\beta_{\rm and}$, and $\beta_{\rm xor}$ as a function of $\beta$ before reaching $\beta_{\rm oax}$.

\mypara{\bf $\beta_{\rm or}$ of $x$-OR-PUF}
Assuming that $i$ out of $x$ PUFs responses flip, the probability of $x$-OR-PUF response $r_{\rm or}$ flips is:

\begin{equation}
\frac{C_{x}^{i}+1}{(x+1)C_{x}^{i}},
\end{equation}
then 

\begin{equation}
\begin{aligned}
\beta_{\rm or}&=\sum_{i=1}^{x}\frac{C_{x}^{i}+1}{(x+1)C_{x}^{i}}(C_{x}^{i}\beta^{i}(1-\beta)^{x-i})\\&=\frac{1}{x+1}[1-(1-\beta )^x+\frac{\beta(\beta^x-(1-\beta)^x)}{2\beta-1}].
\end{aligned}
\end{equation}

\mypara{\bf $\beta_{\rm and}$ of $y$-AND-PUF}

Similarly, $\beta_{\rm and}$ is expressed as:
\begin{equation}
\begin{aligned}
\beta_{\rm and}&=\sum_{i=1}^{y}\frac{C_{y}^{i}+1}{(y+1)C_{y}^{i}}(C_{y}^{i}\beta^{i}(1-\beta)^{y-i})\\&=\frac{1}{y+1}[1-(1-\beta)^y+\frac{\beta(\beta^y-(1-\beta)^y)}{2\beta-1}].
\end{aligned}
\end{equation}

\mypara{\bf $\beta_{\rm xor}$ of $z$-XOR-PUF}

For the $z$-XOR-PUF, if an odd number of PUF responses are flipped, the response $r_{\rm xor}$ will be flipped. This result indicates that $r_{\rm xor}$ will remain unchanged if even number of APUFs responses are flipped. Therefore, $\beta_{\rm xor}$ can be expressed as:

\begin{equation}
\begin{aligned}
\beta_{\rm xor}&=\sum_{i=1,~i\ \text{is odd}}^{z}(C_{z}^{i}\beta^{i}(1-\beta)^{z-i})\\&=\frac{1-(1-2\beta)^z}{2}.
\end{aligned}
\end{equation}

\mypara{\bf $\beta_{\rm oax}$ of ($x,y,z$)-OAX-PUF}

$r_{\rm oax}$ is XORed based on $r_{\rm or}$, $r_{\rm and}$, and $r_{\rm xor}$. It flips when one of $r_{\rm or}$, $r_{\rm and}$, and $r_{\rm xor}$ flips (other two unchanged) or all $r_{\rm or}$, $r_{\rm and}$, and $r_{\rm xor}$ flip. Thus, $\beta_{\rm oax}$ is expressed as:

\begin{equation}
\begin{aligned}
\beta_{\rm oax}&=\beta_{\rm or}\beta_{\rm and}\beta_{\rm xor}\\&+\beta_{\rm or}(1-\beta_{\rm and})(1-\beta_{\rm xor})\\&+\beta_{\rm and}(1-\beta_{\rm or})(1-\beta_{\rm xor})\\&+\beta_{\rm xor}(1-\beta_{\rm or})(1-\beta_{\rm and}).
\end{aligned}
\end{equation}

\subsection{Uniformity}
Uniformity denotes the proportion of `0’ or `1’ out of a PUF's response bits. For truly random PUF responses, the proportion is
50\%~\cite{Maiti11asystematic}. The uniformity is calculated as~\cite{Maiti11asystematic}:
\begin{equation}
\label{eq:u}
    \text{Uniformity} = \frac{1}{n}\sum_{l=1}^{n}r_{l} \times 100\%,
\end{equation}
where $n$ is the number of responses, and the uniformity in this study uses the proportion of `1'.

Below, we use $U_1$ to denote the uniformity with the proportion of `1' used. Again, we start by analyzing the uniformity of $x$-OR-PUF, $y$-AND-PUF, and $z$-XOR-PUF, before reaching the uniformity of ($x,y,z$)-OAX-PUF. The uniformity of each underlying PUF instance is assumed to be the same as 0.5.

\mypara{\bf $U_{\rm or}$ of $x$-OR-PUF}

For the OR logic operation, the output will be `0' only if all $x$ inputs are `0's, therefore, $U_{\rm or_{1}}$ is expressed as:
\begin{equation}
U_{\rm or_{1}}=1-\frac{1}{2^x}.
\end{equation}

We can see that $x$-OR-PUF has a bias towards response `1', especially when $x$ is large.

\mypara{\bf $U_{\rm and}$ of $y$-AND-PUF}

For the AND logic operation, the output will be `1' only if all $y$ inputs are `1's; therefore, $U_{\rm and_{1}}$ is expressed as:

\begin{equation}
U_{\rm and_{1}}=\frac{1}{2^y}.
\end{equation}

We can see that $y$-AND-PUF has a bias towards response `0', especially when $y$ becomes large.

\mypara{\bf $U_{\rm xor}$ of $z$-XOR-PUF}

It is observed that when the response of $z$-XOR-PUF is 1, the number of `1's participating in the XOR logic operation is odd; when the response of $z$-XOR-PUF is `0', the number of `1's participating in XOR operation is even. Therefore, $U_{\rm xor_1}$ can be expressed as:

\begin{equation}
\begin{aligned}
U_{\rm xor_{1}}&=\frac{1}{2^z}(C_{z}^{1}+C_{z}^{3}+...)\\
&=\frac{1}{2^z}{2^{z-1}}\\
&=\frac{1}{2}.
\end{aligned}
\end{equation}

We can see that $z$-XOR-PUF does not have a bias.

\mypara{\bf $U_{\rm oax}$ of ($x,y,z$)-OAX-PUF}

If i) one of $r_{\rm or}$, $r_{\rm and}$, $r_{\rm xor}$ is `1',  and ii) all three equal `1', the response $r_{\rm oax}$ is `1'. Otherwise, the response $r_{\rm oax}$ is `0'. According to $U_{\rm or_{1}}$, $U_{\rm and_{1}}$,and $U_{\rm xor_{1}}$, the uniformity of OAX-PUF is expressed as follows:
\begin{equation}\label{eq:u_oax1}
\begin{aligned}
U_{1}&=U_{\rm or_{1}}U_{\rm and_{1}}U_{\rm xor_{1}}
        +U_{\rm or_{1}}(1-U_{\rm and_{1}})(1-U_{\rm xor_{1}})\\
        &+U_{\rm and_{1}}(1-U_{\rm or_{1}})(1-U_{\rm xor_{1}})\\
        &+U_{\rm xor_{1}}(1-U_{\rm and_{1}})(1-U_{\rm or_{1}})\\
        &=(1-\frac{1}{2^x})\frac{1}{2^y}\frac{1}{2}+(1-\frac{1}{2^x})(1-\frac{1}{2^y})\frac{1}{2}\\
        &+\frac{1}{2^y}\frac{1}{2^x}\frac{1}{2}+\frac{1}{2}(1-\frac{1}{2^y})\frac{1}{2^x}\\
        &=\frac{(2^x-1)+(2^x-1)(2^y-1)}{2^{x+y+1}}\\
        &+\frac{1+(2^y-1)}{2^{x+y+1}}\\
        &=\frac{2^{x+y}}{2^{x+y+1}}\\
        &=\frac{1}{2}.
\end{aligned}
\end{equation}

Thus, we have shown that the uniformity of ($x,y,z$)-OAX-PUF is 0.5 on the condition that the underlying PUF has a uniformity of 0.5. The biased uniformity in the $x$-OR-PUF and $y$-AND-PUF do not propagate and affect the ultimate uniformity of ($x,y,z$)-OAX-PUF. 

\begin{figure}
	\centering
	\includegraphics[trim=0 0 0 0,clip,width=0.5\textwidth]{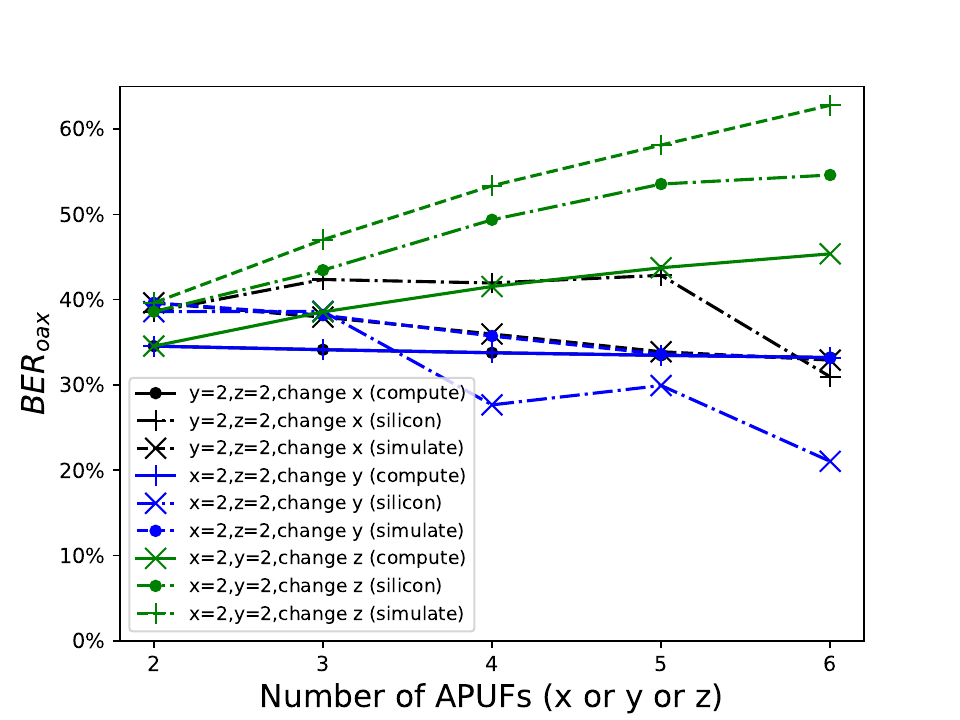}
	\caption{The BER of OAX-PUF (BER$_{\rm oax}$) as a function of a number of PUFs in the OR, AND, and XOR blocks---APUF as an underlying PUF. For the silicon measurement based on RO synthesized APUF, the regenerated response is at (25\textcelsius, 0.96V)---this gives the worst BER---and the enrolled response done at (25\textcelsius, 1.20V) (Virginia Tech Dataset~\cite{maiti2010large}).
	}
	\label{fig:BER}
\end{figure}

\begin{table}[]
\caption{OAX-PUF uniformity validation.}
\label{tab:uniformity}
\centering
\renewcommand{\arraystretch}{1.2}
\begin{tabular}{|c|c|c|c|c|c|c|c|}
\hline
\multirow{2}{*}{Num.APUFs} & \multirow{2}{*}{x} & \multirow{2}{*}{y} & \multirow{2}{*}{z} & OR           & AND           & XOR           & OAX     \\ \cline{5-8} 
                           &                    &                    &                    & U$_{or_{1}}$ & U$_{and_{1}}$ & U$_{xor_{1}}$ & U$_{1}$ \\ \hline
\multirow{3}{*}{5}         & 1                  & 2                  & 2                  & 0.4955       & 0.2500        & 0.5040        & 0.5059  \\ \cline{2-8} 
                           & 2                  & 1                  & 2                  & 0.7484       & 0.4985        & 0.5001        & 0.4984  \\ \cline{2-8} 
                           & 2                  & 2                  & 1                  & 0.7453       & 0.2466        & 0.5000        & 0.5021  \\ \hline
\multirow{4}{*}{6}         & 2                  & 2                  & 2                  & 0.7414       & 0.2459        & 0.4965        & 0.5092  \\ \cline{2-8} 
                           & 1                  & 3                  & 2                  & 0.4960       & 0.1221        & 0.5082        & 0.4929  \\ \cline{2-8} 
                           & 3                  & 1                  & 2                  & 0.8752       & 0.4957        & 0.5060        & 0.5007  \\ \cline{2-8} 
                           & 2                  & 3                  & 1                  & 0.7520       & 0.1196        & 0.5060        & 0.4994  \\ \hline
\multirow{4}{*}{7}         & 3                  & 2                  & 2                  & 0.8786       & 0.2484        & 0.5034        & 0.4964  \\ \cline{2-8} 
                           & 2                  & 3                  & 2                  & 0.7545       & 0.1268        & 0.4975        & 0.5086  \\ \cline{2-8} 
                           & 2                  & 2                  & 3                  & 0.7530       & 0.2504        & 0.5016        & 0.5002  \\ \cline{2-8} 
                           & 2                  & 4                  & 1                  & 0.7592       & 0.0634        & 0.5035        & 0.4985  \\ \hline
\multirow{4}{*}{8}         & 4                  & 2                  & 2                  & 0.9398       & 0.2477        & 0.4992        & 0.5019  \\ \cline{2-8} 
                           & 2                  & 4                  & 2                  & 0.7405       & 0.0618        & 0.5015        & 0.4968  \\ \cline{2-8} 
                           & 2                  & 2                  & 4                  & 0.7418       & 0.2465        & 0.5004        & 0.4989  \\ \cline{2-8} 
                           & 2                  & 5                  & 1                  & 0.7478       & 0.0315        & 0.4941        & 0.5024  \\ \hline
\multirow{4}{*}{9}         & 5                  & 2                  & 2                  & 0.9664       & 0.2490        & 0.5012        & 0.5008  \\ \cline{2-8} 
                           & 2                  & 5                  & 2                  & 0.7522       & 0.0348        & 0.4935        & 0.4965  \\ \cline{2-8} 
                           & 2                  & 2                  & 5                  & 0.7578       & 0.2468        & 0.5086        & 0.5042  \\ \cline{2-8} 
                           & 2                  & 6                  & 1                  & 0.7423       & 0.0144        & 0.5053        & 0.4994  \\ \hline
\end{tabular}
\end{table}

\subsection{Validation}

\noindent{\bf Reliability:} First, we validate the correctness of the proposed derived reliability formulation. Second, we show that APUF in which block---OR, AND, and XOR---has a higher impact on the unreliability or BER of OAX-PUF.

There are three parameters: $x,y,z$. We fix two of them to be 2 and change the remaining one from 2 to 6, and the results are shown in Fig.~\ref{fig:BER}. For the simulation, the BER of each APUF is approximately 13\% and assigns a small variance for each PUF instance. For the computation based on the proposed formulation, the BER of each APUF is set to be 13\% with no variance. For the silicon measurement, we build OAX-APUF$_{RO}$ with RO synthesized APUFs with the Virginia Tech dataset~\cite{maiti2010large}. For the worst-case BER of 13\%, the response is regenerated at operating corner of (25\textcelsius, 0.96V), while the enrollment operating corner is (25\textcelsius, 1.20V). More details of the silicon measurement settings are provided in the Appendix. Notably, the 13\% BER per APUF for the simulation and formulation  essentially aligns with the BER of the silicon measurements for fair comparison.
First, both the simulated and silicon results match  the proposed formulation with the same tendency, validating the formulation's correctness. Second, we can see that increasing $x$ and $y$---the number of PUFs in the OR and AND block, respectively---does not increase the BER of OAX-PUF. In contrast, increasing $z$---the number of PUFs in the XOR block---increases $\beta_{\rm oax}$, which means that we have successfully disrupted the underlying PUF's unreliability contribution to the recomposited OAX-PUF's unreliability. More specifically, the PUFs in the XOR block have higher contribution while those in the OR and AND blocks contribute less. From the security perspective, this prevents reliability based CMA-ES attacks from finding PUFs components in the OR and AND blocks, as experimentally validated in Section~\ref{sec:CMAESattack}.

\mypara{Uniformity}
We now simulate $x+y+z$ APUF vectors, where each vector has $10,000$ elements of `1' or `0'---the `1'/`0' is randomly generated with a probability of 0.5, assuming that each underlying PUF is with an ideal uniformity of 0.5. Then, the first $x$ PUF vector(s) are ORed to resemble $U_{\rm or}$. The next $y$ vectors are ANDed to resemble $U_{\rm and}$, and the remaining $z$ vector(s) are OXRed to resemble $U_{\rm xor}$. Finally, the three new vectors from the OR, AND, and XOR blocks are XORed to resemble the ultimate uniformity of OAX-PUF. Table~\ref{tab:uniformity} summarizes the simulated results when a number of $x,y,z$ combinations are used, where the uniformity of OAX-PUF always approaches 0.5. This result agrees well with the formulation in Eq.~\ref{eq:u_oax1}, again validating the effectiveness of the proposed formulation.

\section{Resilience Against Modeling Attacks}\label{Sec:securityEvaluations}

To thoroughly examine to what extent the OAX-PUF can protect against modeling attacks, \textit{we use the APUF as a basic PUF instance to form the OAX-APUF due to its popularity as an empirical study}. We also compare OAX-APUF with XOR-APUF, which is also a well-studied APUF variant under modeling attacks. Analysis of each modeling attack result is provided.

Following~\cite{ruhrmair2010modeling,becker2015pitfalls,nguyen2019interpose,wisiol2020splitting}, we use the standard means to simulate the APUF, which has been recognized as an efficient and common way when evaluating the performance of APUF or its variants~\cite{ruhrmair2013puf,nguyen2019interpose}. We assume that the weight elements in the weight vector $\vec{w}$ follow a normal distribution with $\mu = 0$ and $\sigma= 1$. The noise for each weight follows a normal distribution $ N (0, \sigma^{2}_{\rm noise})$, $\sigma_{\rm noise}=0.05$. The distribution of each weight is therefore $N (0, \sigma^{2} + \sigma_{\rm noise}^{2})$.
% CRP

For CRPs, we first randomly generate the required number of 64 bit challenges, which are composed of `0's and `1's, and calculate the feature vector of each challenge according to Eq. \ref{con:feature vector}. Then, the $\Delta$ of each APUF is calculated according to Eq. \ref{con:delta}, and the corresponding response is obtained according to Eq. \ref{con:responses}. Finally, the response of each APUF is processed to tobtain the final response of XOR-APUF or OAX-APUF according to their corresponding logic operations.

We use a Python simulator$\footnote{The source code is based on \url{https://github.com/nils-wisiol/puf-lr/blob/master/lr.py } released by Nils Wisiol.}$ to generate the CRPs required for LR and MLP attacks.
For the CMA-ES attack, we use a MATLAB simulator$\footnote{The source code is based on \url{https://github.com/scluconn/DA\_PUF\_Library/tree/master/MatLab\_simulation/Becker\_Attack}~\cite{nguyen2019interpose}}$ to generate CRPs. The LR and MLP attacks  use the Python simulator because it can complete attacks on large-scaled OAX-APUF or XOR-APUF in our setting. Due to insufficient memory, the MATLAB simulator cannot generate $20,000,000$ training CRPs. For the hybrid LR-reliability attack, we add the implementation of OAX based on the code$\footnote{\url{https://github.com/jtobi/puf-simulation}}$~\cite{tobisch2021combining}.

\subsection{Evaluating XOR-APUF and OAX-APUF against LR}\label{sec:evaluateMLP}

\begin{table*}[]
\caption{LR attack on XOR-APUF(($x=0,y=0,z=l$)-OAX-APUF) and ($x,y,z$)-OAX-APUF.}
\centering
\label{tab1}
\resizebox{\textwidth}{!}{
\begin{threeparttable}
\renewcommand{\arraystretch}{1.2}
\begin{tabular}{|c|c|c|c|c|c|c|c|c|c|}
\hline
\textbf{\begin{tabular}[c]{@{}c@{}}Num.\\ APUFs\end{tabular}} & \textbf{\begin{tabular}[c]{@{}c@{}}Training\\  CRPs\end{tabular}} & \textbf{\begin{tabular}[c]{@{}c@{}}Test \\ CRPs\end{tabular}} & \textbf{Epochs}     & \textbf{Batch\_size}   & \textbf{x,y,z} & \textbf{Best Pred.Acc} & \textbf{Training Time} & \textbf{CPU}                                                                                                       & \textbf{RAM}                                                                             \\ \hline
\multirow{4}{*}{5}                                            & \multirow{4}{*}{200,000}                                          & \multirow{4}{*}{15,000}                                        & \multirow{4}{*}{20} & \multirow{4}{*}{1,000} & 1,2,2          & 94.10\%                & 9.5 min                & \multirow{9}{*}{\begin{tabular}[c]{@{}c@{}}Intel(R) Core(TM) \\ i5-6200U CPU \end{tabular}} & \multirow{9}{*}{\begin{tabular}[c]{@{}c@{}}12.0 GB\end{tabular}} \\ \cline{6-8}
                                                              &                                                                   &                                                               &                     &                        & 2,1,2          & 92.70\%                & 9.03 min               &                                                                                                                    &                                                                                          \\ \cline{6-8}
                                                              &                                                                   &                                                               &                     &                        & 2,2,1          & 93.69\%                & 18.85 min              &                                                                                                                    &                                                                                          \\ \cline{6-8}
                                                              &                                                                   &                                                               &                     &                        & 0,0,5          & 93.50\%                & 5.31 s                 &                                                                                                                    &                                                                                          \\ \cline{1-8}
\multirow{5}{*}{6}                                            & \multirow{5}{*}{1,400,000}                                        & \multirow{5}{*}{15,000}                                        & \multirow{5}{*}{20} & \multirow{5}{*}{1,000} & 2,2,2          & 99.09\%                & 56.6 min               &                                                                                                                    &                                                                                          \\ \cline{6-8}
                                                              &                                                                   &                                                               &                     &                        & 1,3,2          & 89.99\%                & 42.95 min              &                                                                                                                    &                                                                                          \\ \cline{6-8}
                                                              &                                                                   &                                                               &                     &                        & 3,1,2          & 88.30\%                & 40.64 min              &                                                                                                                    &                                                                                          \\ \cline{6-8}
                                                              &                                                                   &                                                               &                     &                        & 2,3,1          & 90.20\%                & 1.17 h                 &                                                                                                                    &                                                                                          \\ \cline{6-8}
                                                              &                                                                   &                                                               &                     &                        & 0,0,6          & 98.00\%                & 30.62 s                &                                                                                                                    &                                                                                          \\ \hline
\multirow{5}{*}{7}                                            & \multirow{5}{*}{20,000,000}                                       & \multirow{5}{*}{15,000}                                        & \multirow{5}{*}{20} & \multirow{5}{*}{500}   & 3,2,2          & 78.10\%                & 12.29 h                & \multirow{5}{*}{\begin{tabular}[c]{@{}c@{}}Intel(R) Core(TM)\\  i7-9750H CPU \end{tabular}} & \multirow{5}{*}{\begin{tabular}[c]{@{}c@{}}16.0 GB\end{tabular}} \\ \cline{6-8} 
                                                          
                                                              &                                                                   &                                                               &                     &                        & 2,3,2          & 89.30\%                & 10.99 h                &                                                                                                                    &                                                                                          \\ \cline{6-8} 
                                                              &                                                                   &                                                               &                     &                        & 2,2,3          & 72.30\%                & 9.44 h                 &                                                                                                                    &                                                                                          \\ \cline{6-8} 
                                                              &                                                                   &                                                               &                     &                        & 2,4,1          & 80.00\%                & 14.36 h                &                                                                                                                    &                                                                                          \\ \cline{6-8} 
                                                              &                                                                   &                                                               &                     &                        & 0,0,7          & 97.10\%                & 9.94 min               &                                                                                                                    &                                                                                           \\ \hline
\end{tabular}

\begin{tablenotes}
        \footnotesize
        \item[*] The ($1,2,2$)-OAX-APUF can be understood as exactly equal to ($0,2,3$)-OAX-APUF, in which the XOR block is composed of APUF1, APUF4 and APUF5. ($2,1,2$)-OAX-APUF can be understood as ($2,0,3$)-OAX-APUF, in which the XOR block is composed of APUF3, APUF4 and APUF5. In these cases, OR, or AND logic operation is meaningless. Therefore, if there is only 1 OR or AND in the combination, the number of APUFs participating in XOR operation is actually increased by 1, and the OR or AND logic operation can be removed. 
      \end{tablenotes}
    \end{threeparttable}
    }
\end{table*}

The objective of this experiment is to first reproduce LR attack on $l$-XOR-APUF and then use LR to attack ($x,y,z$)-OAX-APUF for end-to-end modeling resilience comparisons, with $l=x+y+z$ same number of APUF instances or area overhead.

\mypara{Setup}
We first run LR attacks$ \footnote{LR is implemented by keras, which does not provide RPROP optimizer; thus,we use the RMSProp optimizer and small batch training. RMSProp is an improved version of RPROP algorithm.}$ on $l$-XOR-APUF where $l=5,6,7$. In all our experiments, the APUF has 64 stages. The CRPs are randomly generated. A higher $l$ usually requires a larger number of CRPs for model training to gain a high prediction accuracy. Following~\cite{tobisch2015scaling,nguyen2019interpose}, the number of CRPs used when $l=5$, $l=6$ and $l=7$ are $2 \times 10^{5}$, $1.4 \times 10^{6}$ and $20 \times 10^{6}$, respectively. With regard to epochs and batch sizes, most are set to $20$ and $1,000$, respectively.  However, $x+y+z=7$ is special due to the limitations of computer performance, the epoch is $20$, and the batch size is $500$. This work does not reproduce LR attacks on $l$-XOR-APUF for $l=8$ and $l=9$ because this attack is extremely time and resource hungry beyond the capability of the authors’ computing resources. For example, the optimal number of CRPs used for LR attack when $l=8$ and $l=9$ are up to $150\times 10^{6}$ and $350\times 10^{6}$, respectively. Thus, the former requires more than 6 hours and 12.3 GB memory, the latter requires 37 hours and 31.5 GB memory using an AMD Opteron cluster, which consists of 4 nodes, each with 64 cores and 256 GB of memory---though each run uses 16 cores~\cite{tobisch2015scaling}. It is clear that such requirements are beyond the capability of a medium-end PC with an Intel(R) Core(TM) i5-6200U CPU, and 12~GB memory. \textit{This PC is used for most experiments unless otherwise stated}. Note that for the case of $x+y+z=7$, we use 
a different medium-end PC with an Intel(R) Core(TM) i7-9750H CPU and 16~GB memory to run an LR attack at a different time.

For the LR attacks on $(x,y,z)$-OAX-APUF, we have attacked various $x,y,z$ parameter combinations while keeping $x+y+z = l$ to ensure that the same number of APUF components are set for end-to-end comparisons. In addition, the number of CRPs used for model training is also kept the same. For all combinations of LR attacks, we used $15,000$ CRPs for model prediction during the testing phase to evaluate the attack accuracy. 

\mypara{Analysing Results} 

The results of the LR attack on $l$-XOR-APUFs and ($x,y,z$)-OAX-APUFs are summarized in Table~\ref{tab1}. The best prediction accuracy of LR attack on the $l$-XOR-APUF is above 97\% for $l=6,7$. When $l = 5$, the best prediction accuracy also reaches 93.5\%. The larger $l$ is, the longer the time the attack requires for model training. For OAX-APUF, when $x+y+z=5$, the best prediction accuracy of the LR attack is similar to that of $5$-XOR-APUF and even exceeds that of $5$-XOR-APUF when $x = 1, y = 2,z = 2$. When $x = 2, y = 2,z = 2$, the accuracy of LR modeling reaches 99\%, which is higher than that of $6$-XOR-APUF. In other cases where $x+y+z=6$, the best prediction accuracy of LR is close to 90\%. When $x+y+z=7$, the accuracy of the LR attack on OAX-APUFs is lower than that of XOR-APUF, and the OAX-APUF seems more difficult to break. Generally, OAX-APUF exhibits improved modeling resilience, especially for slightly large-scale cases (e.g., $x+y+z$ is 7). The time to attack the OAX-APUF is  also substantially prolonged. 

\subsection{Evaluating XOR-APUF and OAX-APUF against CMA-ES}\label{sec:CMAESattack}
Similarly, we first reproduce the reliability based CMA-ES attack on $l$-XOR-APUFs and then mount it on ($x,y,z$)-OAX-APUFs for an end-to-end comparison.

\mypara{Setup}
We run the CMA-ES attack on the same set of APUFs for $l$-XOR-APUF and ($x,y,z$)-OAX-APUF. We implemented 5, 6, 7, 8, 9 APUFs. The reliability of a silicon APUF is typically more than 90\%, and we set $\sigma_{noise}$ to 0.05, which yields an APUF BER ranging from 5.5\% to 8.5\% in most of the simulations---this is aligned with the fabricated ASIC APUFs BER of 5.89\% in~\cite{maes2013physically}. The CMA-ES attack stops when one of the following two conditions is satisfied: the number of CMA-ES iterations reaches $30,000$; the fitness (i.e., error) is sufficient (e.g., less than $10^{-10}$). The number of challenge reliability pairs used for training given different sets of numbers of APUFs in $l$-XOR-APUF and ($x,y,z$)-OAX-APUF are in Table~\ref{tab:CMA-ES}. The number of CRPs used for model prediction accuracy is $15,000$. To determine the short-time reliability (according to Eq.~\ref{eq:shortReliability}) of a response, the same challenge is repeatedly queried 11 times to the same PUF instance.

\mypara{Analysing Results} 
The results of the CMA-ES attack on $l$-XOR-APUFs and ($x,y,z$)-OAX-APUFs are summarized in Table~\ref{tab:CMA-ES}. When the accuracy of test CRPs is more than 90\% or less than 10\%, the APUF is considered to be converged. Note here we repeatedly run the CMA-ES attack on the $l$-XOR-APUF and ($x,y,z$)-OAX-APUF both $l$ times, and $l=x+y+z$. It can be seen that the number of converged APUFs underling the $l$-XOR-APUF is markedly higher than that of ($x,y,z$)-OAX-APUF. As expected, as for the $l$-XOR-APUF itself, the converged APUFs always have higher BERs. However, because the APUF unreliability contribution to the XOR-APUF is not particularly distinguishable, the CMA-ES attack can find more underling APUFs. Kindly note that once $x=1$ or $y=1$, the OR block and AND block are meaningless and equally join the XOR block---number of APUFs in XOR block increases by one. Now, we see that the converged APUFs are all located on those in the XOR block in the OAX-APUF case, preventing convergence to APUFs in the OR and AND blocks.
The unreliability contribution of APUFs in the XOR block is dominant in OAX-APUF, as we have formulated and validated in Section~\ref{sec:oax-apuf}, which is essentially the impetus of devising OAX-APUF.

\subsection{Evaluating XOR-APUF and OAX-APUF against hybrid LR-reliability attack}

Similar to the above attacks, the purpose of this experiment is to check to what extent OAX-APUF can resist hybrid LR-reliability attacks or whether it has any advantages compared with the resistance of XOR-APUF. The number of CRPs required for the success of hybrid LR-reliability modeling attacks is the smallest~\cite{tobisch2021combining}.

\mypara{Setup}
The hybrid LR-reliability attack simultaneously uses challenge response pairs and challenge reliability pairs. To obtain the reliability information of the response, we set the repeated measurement of the same challenge 10 times to be consistent with~\cite{tobisch2021combining}. For batch size and epoch number, the settings are $256$ and $25$, respectively. It is worth noting that the attack can train multiple models concurrently, and the final prediction accuracy is the best prediction accuracy. For each combination, 6 models were trained at the same time. Regarding the number of CRPs required for training, for XOR-APUF, most of them are set to be consistent with~\cite{tobisch2021combining}. They are $30,000$ when $l = 5$, $40,000$ when $l = 6$, $60,000$ when $l = 7$, $80,000$ when $l = 8$ and $90,000$ when $l = 9$. We find that OAX-APUF needs marginally more CRPs.
For $\epsilon_{1}^{xor}$ and $\epsilon_{2}^{xor}$ and $\epsilon_{3}^{xor}$ in the loss function, the experiment selects the same parameter settings as ~\cite{tobisch2021combining}: $\epsilon_{1}^{xor} = 12$, $\epsilon_{2}^{xor}= 1$, and $\epsilon_{b}^{xor} = 0.2$.

\mypara{Analysing Results}
The results of the hybrid LR-reliability attack are summarized in Table~\ref{combined-Attack}. We can see that the best prediction accuracy of this attack on XOR-APUFs is primarily more than 90\%. For OAX-APUF, when $x + y + z\leq 7$, the number of CRPs required for successful modeling is greater than XOR-APUF. When $x + y + z = 8$, the best prediction accuracy of some OAX-APUFs can reach more than 90\%. The accuracy of the other cases is approximately 90\%.  Interestingly, when $x + y + z = 9$, XOR-APUF requires more CRPs than OAX-APUFs selected. The best prediction accuracy of these OAX-APUFs is more than 90\%.

From these experimental results, the attacking accuracy of OAX-APUF against hybrid LR-reliability attack is similar to that of XOR-APUF. The reason behind this attack is that the third term of the loss function used by this attack restricts the possibility of converging to the same APUF. Therefore, although each APUF in OAX-APUF has a different contribution to the reliability of the last output, it can still be broken by this attack opposed to the purely reliability based CMA-ES attack. Conversely, the time required to model OAX-APUF is markedly higher than that required to model XOR-APUF. 

\begin{table*}[]
\caption{Reliability enabed CMA-ES attack on XOR-APUF(($x=0,y=0,z=l$)-OAX-APUF) and ($x,y,z$)-OAX-APUF.}
\label{tab:CMA-ES}
\resizebox{\textwidth}{!}{
\begin{threeparttable}
\centering
\renewcommand{\arraystretch}{1.2}
\begin{tabular}{|c|c|c|c|c|c|c|c|c|c|c|c|c|c|}
\hline
\multirow{2}{*}{\textbf{PUFs}} & \multirow{2}{*}{\textbf{x,y,z}} & \multirow{2}{*}{\textbf{\begin{tabular}[c]{@{}c@{}}Training\\ CRPs\end{tabular}}} & \multirow{2}{*}{\textbf{\begin{tabular}[c]{@{}c@{}}Test\\ CRPs\end{tabular}}} & \multicolumn{9}{c|}{\textbf{\begin{tabular}[c]{@{}c@{}}BER\_APUFs\\ (BER measurement: generate 10,000 CRPs randomly for each operation)\end{tabular}}}                                       & \multirow{2}{*}{\textbf{BER$_{\rm oax}$}} \\ \cline{5-13}
                               &                                 &                                                                                   &                                                                               & \textbf{APUF1}     & \textbf{APUF2}    & \textbf{APUF3}     & \textbf{APUF4}     & \textbf{APUF5}     & \textbf{APUF6}     & \textbf{APUF7}     & \textbf{APUF8}     & \textbf{APUF9}     &                                    \\ \hline
\multirow{3}{*}{5(OAX)}        & 1,2,2                           & \multirow{4}{*}{260,000}                                                          & \multirow{4}{*}{15,000}                                                          & \textbf{0.0669(5)} & 0.0539            & 0.0575             & 0.0549             & 0.0559             &                    &                    &                    &                    & 0.2161                             \\ \cline{2-2} \cline{5-14} 
                               & 2,1,2                           &                                                                                   &                                                                               & 0.0688             & 0.0588            & 0.0643             & \textbf{0.0576(3)} & \textbf{0.0602(1)} &                    &                    &                    &                    & 0.2201                             \\ \cline{2-2} \cline{5-14} 
                               & 2,2,1                           &                                                                                   &                                                                               & 0.0721             & 0.0572            & 0.0587             & 0.0619             & \textbf{0.0569(1)} &                    &                    &                    &                    & \textbf{0.1623}                    \\ \cline{1-2} \cline{5-14} 
5 XOR                          & 0,0,5                           &                                                                                   &                                                                               & \textbf{0.0658(3)} & 0.0573            & 0.0614             & \textbf{0.0608(1)} & \textbf{0.0612(1)} &                    &                    &                    &                    & \textbf{0.2701}                    \\ \hline
\multirow{4}{*}{6(OAX)}        & 2,2,2                           & \multirow{5}{*}{600,000}                                                          & \multirow{5}{*}{15,000}                                                          & 0.0645             & 0.0711            & 0.0528             & 0.0566             & \textbf{0.0642(3)} & 0.0664             &                    &                    &                    & 0.2267                             \\ \cline{2-2} \cline{5-14} 
                               & 1,3,2                           &                                                                                   &                                                                               & \textbf{0.065(1)}  & 0.0633            & 0.0527             & 0.053              & \textbf{0.062(2)}  & \textbf{0.0626(3)} &                    &                    &                    & 0.2059                             \\ \cline{2-2} \cline{5-14} 
                               & 3,1,2                           &                                                                                   &                                                                               & 0.0685             & 0.0667            & 0.0561             & \textbf{0.0513(1)} & \textbf{0.0645(4)} & \textbf{0.0585(1)} &                    &                    &                    & 0.2033                             \\ \cline{2-2} \cline{5-14} 
                               & 2,3,1                           &                                                                                   &                                                                               & 0.064              & 0.0662            & 0.0548             & 0.0562             & 0.0616             & \textbf{0.0614(1)} &                    &                    &                    & \textbf{0.1608}                    \\ \cline{1-2} \cline{5-14} 
6XOR                           & 0,0,6                           &                                                                                   &                                                                               & \textbf{0.0634(1)} & \textbf{0.065(1)} & 0.0534             & \textbf{0.0543(1)} & \textbf{0.0581(3)} & 0.059              & \textbf{}          & \textbf{}          & \textbf{}          & \textbf{0.3054}                    \\ \hline
\multirow{4}{*}{7(OAX)}        & 3,2,2                           & \multirow{5}{*}{700,000}                                                          & \multirow{5}{*}{15,000}                                                          & 0.0622             & 0.0609            & 0.0495             & 0.0661             & 0.0598             & \textbf{0.0674(6)} & \textbf{0.0581(1)} &                    &                    & 0.2151                             \\ \cline{2-2} \cline{5-14} 
                               & 2,3,2                           &                                                                                   &                                                                               & 0.0639             & 0.0607            & 0.05               & 0.0548             & 0.0646             & \textbf{0.0741(4)} & \textbf{0.0516(1)} &                    &                    & 0.2181                             \\ \cline{2-2} \cline{5-14} 
                               & 2,2,3                           &                                                                                   &                                                                               & 0.0673             & 0.0573            & 0.0524             & 0.065              & 0.063              & \textbf{0.0656(4)} & 0.0618             &                    &                    & 0.2824                             \\ \cline{2-2} \cline{5-14} 
                               & 2,4,1                           &                                                                                   &                                                                               & 0.0722             & 0.0638            & 0.0556             & 0.0621             & 0.0595             & 0.0676             & \textbf{0.0588(4)} &                    &                    & \textbf{0.1561}                    \\ \cline{1-2} \cline{5-14} 
7 XOR                          & 0,0,7                           &                                                                                   &                                                                               & \textbf{0.0715(2)} & 0.0614            & \textbf{0.0517(1)} & \textbf{0.0674(1)} & 0.063              & \textbf{0.0699(3)} & 0.0585             & \textbf{}          & \textbf{}          & \textbf{0.3681}                    \\ \hline
\multirow{4}{*}{8(OAX)}        & 4,2,2                           & \multirow{5}{*}{900,000}                                                          & \multirow{5}{*}{15,000}                                                          & 0.0583             & 0.0546            & 0.0544             & 0.0586             & 0.06               & 0.0631             & \textbf{0.0787(3)} & \textbf{0.0817(5)} & \textbf{}          & 0.225                              \\ \cline{2-2} \cline{5-14} 
                               & 2,4,2                           &                                                                                   &                                                                               & 0.0574             & 0.0599            & 0.0527             & 0.0572             & 0.0553             & 0.0617             & \textbf{0.0806(5)} & \textbf{0.0851(2)} & \textbf{}          & 0.2258                             \\ \cline{2-2} \cline{5-14} 
                               & 2,2,4                           &                                                                                   &                                                                               & 0.0586             & 0.0663            & 0.056              & 0.0569             & \textbf{0.0611(1)} & \textbf{0.0629(1)} & \textbf{0.0736(5)} & 0.0763             & \textbf{}          & 0.3289                             \\ \cline{2-2} \cline{5-14} 
                               & 2,5,1                           &                                                                                   &                                                                               & 0.0588             & 0.0582            & 0.0521             & 0.0586             & 0.059              & 0.061              & 0.0768             & \textbf{0.0753(8)} & \textbf{}          & \textbf{0.1357}                    \\ \cline{1-2} \cline{5-14} 
8 XOR                          & 0,0,8                           &                                                                                   &                                                                               & 0.0541             & 0.0595            & 0.0512             & 0.0603             & 0.069              & 0.0607             & \textbf{0.0848(6)} & \textbf{0.0856(2)} & \textbf{}          & \textbf{0.4212}                    \\ \hline
\multirow{4}{*}{9(OAX)}        & 5,2,2                           & \multirow{5}{*}{1,000,000}                                                        & \multirow{5}{*}{15,000}                                                          & 0.0689             & 0.0649            & 0.058              & 0.0581             & 0.0551             & 0.0631             & 0.0706             & \textbf{0.0725(7)} & \textbf{0.0616(2)} & 0.2029                             \\ \cline{2-2} \cline{5-14} 
                               & 2,5,2                           &                                                                                   &                                                                               & 0.0668             & 0.0705            & 0.055              & 0.0537             & 0.0612             & 0.0666             & 0.0656             & \textbf{0.0693(9)} & 0.0591             & \textbf{0.1848}                    \\ \cline{2-2} \cline{5-14} 
                               & 2,2,5                           &                                                                                   &                                                                               & 0.0657             & 0.0681            & 0.0546             & 0.0523             & 0.0564             & \textbf{0.0613(1)} & 0.0715             & \textbf{0.072(8)}  & 0.0622             & 0.3588                             \\ \cline{2-2} \cline{5-14} 
                               & 2,6,1                           &                                                                                   &                                                                               & 0.0648             & 0.0611            & 0.0575             & 0.0565             & 0.0582             & 0.0654             & 0.0742             & 0.071              & \textbf{0.0624(6)} & \textbf{0.119}                     \\ \cline{1-2} \cline{5-14} 
9XOR                           & 0,0,9                           &                                                                                   &                                                                               & \textbf{0.0627(3)} & 0.0678            & 0.0626             & 0.0522             & 0.0615             & 0.0609             & 0.0673             & \textbf{0.0643(4)} & \textbf{0.0587(2)} & \textbf{0.4355}                    \\ \hline
\end{tabular}
\begin{tablenotes}
        \footnotesize
        \item[*] The training CRPs in the third column refer to the training challenge reliability pairs instead of challenge response pairs.
        \item[**] Ideally, for XOR-APUF, each run converges to a different APUF with the same probability. Adding () after the BER of an APUF instance indicates that CMA-ES attack converges to this APUF instance, and the number in () represents the converged times.
        \item[***] ($1,2,2$)-OAX-APUF can be understood as ($0,2,3$)-OAX-APUF, in which the XOR block is composed of APUF1, APUF4 and APUF5. Based on this, we can find that CMA-ES attacks tend to converge to APUFs with relatively large BER in the \textit{XOR block}.
      \end{tablenotes}
      \end{threeparttable}
      }
\end{table*}

\begin{table*}[]
\caption{Hybrid LR-reliability attack on XOR-APUF(($x=0,y=0,z=l$)-OAX-APUF) and ($x,y,z$)-OAX-APUF.}
\centering
\label{combined-Attack}
\begin{threeparttable}
\renewcommand{\arraystretch}{1.2}
\begin{tabular}{|c|c|c|c|c|c|c|c|c|}
\hline
\textbf{Num.APUFs} & \textbf{x,y,z}         & \textbf{Training CRPs} & \textbf{Test CRPs} & \textbf{Epochs} & \textbf{Batch\_size} & \textbf{Trials} & \textbf{Best Pred.Acc} & \textbf{Training Time} \\ \hline
\multirow{7}{*}{5} & \multirow{2}{*}{1,2,2} & 30,000                 & 15,000              & 25              & 256                  & 6               & 86.75\%                & 1.07h                  \\ \cline{3-9} 
                   &                        & 50,000                 & 15,000              & 25              & 256                  & 6               & 96.40\%                & 1.21h                  \\ \cline{2-9} 
                   & \multirow{2}{*}{2,1,2} & 30,000                 & 15,000             & 25              & 256                  & 6               & 87.30\%                & 1.08h                  \\ \cline{3-9} 
                   &                        & 50,000                 & 15,000              & 25              & 256                  & 6               & 96.70\%                & 2.07h                  \\ \cline{2-9} 
                   & \multirow{2}{*}{2,2,1} & 30,000                 & 15,000              & 25              & 256                  & 6               & 88.60\%                & 1.24h                  \\ \cline{3-9} 
                   &                        & 50,000                 & 15,000              & 25              & 256                  & 6               & 90.00\%                & 2.08h                  \\ \cline{2-9} 
                   & 0,0,5                  & 30,000                 & 15,000              & 25              & 256                  & 6               & 93.40\%                & 2min                   \\ \hline
\multirow{9}{*}{6} & \multirow{2}{*}{2,2,2} & 40,000                 & 15,000              & 25              & 256                  & 6               & 89.80\%                & 2.39h                  \\ \cline{3-9} 
                   &                        & 70,000                 & 15,000              & 25              & 256                  & 6               & 90.65\%                & 3.31h                  \\ \cline{2-9} 
                   & \multirow{2}{*}{1,3,2} & 40,000                 & 15,000              & 25              & 256                  & 6               & 87.65\%                & 2.33h                  \\ \cline{3-9} 
                   &                        & 80,000                 & 15,000              & 25              & 256                  & 6               & 90.10\%                & 2.93h                  \\ \cline{2-9} 
                   & \multirow{2}{*}{3,1,2} & 40,000                 & 15,000              & 25              & 256                  & 6               & 87.00\%                & 2.4h                   \\ \cline{3-9} 
                   &                        & 80,000                 & 15,000              & 25              & 256                  & 6               &89.95\%                 &2.69h                   \\ \cline{2-9} 
                   & \multirow{2}{*}{2,3,1} & 40,000                 & 15,000              & 25              & 256                  & 6               & 87.15\%                & 2.57h                  \\ \cline{3-9} 
                   &                        & 60,000                 & 15,000              & 25              & 256                  & 6               & 91.20\%                & 4h                     \\ \cline{2-9} 
                   & 0,0,6                  & 40,000                 & 15,000              & 25              & 256                  & 6               & 92.25\%                & 2.3min                 \\ \hline
\multirow{5}{*}{7} & 3,2,2                  & 60,000                 & 15,000              & 25              & 256                  & 6               & 90.75\%                & 4.48h                  \\ \cline{2-9} 
                   & 2,3,2                  & 60,000                 & 15,000              & 25              & 256                  & 6               & 88.10\%                & 4.45h                  \\ \cline{2-9} 
                   & 2,2,3                  & 60,000                 & 15,000              & 25              & 256                  & 6               & 87.75\%                & 4.06h                  \\ \cline{2-9} 
                   & 2,4,1                  & 60,000                 & 15,000              & 25              & 256                  & 6               & 89.35\%                & 4.74h                  \\ \cline{2-9} 
                   & 0,0,7                  & 60,000                 & 15,000              & 25              & 256                  & 6               & 91.60\%                & 4.04min                \\ \hline
\multirow{5}{*}{8} & 4,2,2                  & 80,000                 & 15,000              & 25              & 256                  & 6               & 91.40\%                & 6.91h                  \\ \cline{2-9} 
                   & 2,4,2                  & 80,000                 & 15,000              & 25              & 256                  & 6               & 89.40\%                & 6.77h                  \\ \cline{2-9} 
                   & 2,2,4                  & 80,000                 & 15,000              & 25              & 256                  & 6               & 88.40\%                & 5.66h                  \\ \cline{2-9} 
                   & 2,5,1                  & 80,000                 & 15,000              & 25              & 256                  & 6               & 94.45\%                & 7.02h                  \\ \cline{2-9} 
                   & 0,0,8                  & 80,000                 & 15,000              & 25              & 256                  & 6               & 90.20\%                & 4.89min                \\ \hline
\multirow{5}{*}{9} & 5,2,2                  & 90,000                 & 15,000              & 25              & 256                  & 6               & 92.40\%                & 10.58h                 \\ \cline{2-9} 
                   & 2,5,2                  & 90,000                 & 15,000              & 25              & 256                  & 6               & 94.30\%                & 10.28h                 \\ \cline{2-9} 
                   & 2,2,5                  & 90,000                 & 15,000               & 25              & 256                  & 6               & 90.75\%                & 8.82h                  \\ \cline{2-9} 
                   & 2,6,1                  & 90,000                 & 15,000               & 25              & 256                  & 6               & 95.10\%                & 10.75h                 \\ \cline{2-9} 
                   & 0,0,9                  & 90,000                 & 15,000              & 25              & 256                  & 6               & 88.15\%                & 5.62min                \\ \hline
\end{tabular}
\begin{tablenotes}
        \footnotesize
        % \item[*] The numbers before OR, AND, XOR represent the number of APUFs participating in the calculation.
        \item[*] For the combination of 5 APUFs, less CRPs are required for successful modeling of 5-XOR-APUF than OAX-APUF. The same is true for a combination of 6 APUFs. It is speculated that the final modeling accuracy of 7 and 8 APUFs combinations can reach more than 90\%. For OAX-APUFs, the modeling time is higher than XOR-APUFs, and the time is mainly consumed in the calculation of hypothetical function.
\end{tablenotes}
    \end{threeparttable}
\end{table*}

\subsection{Evaluating XOR-APUF and OAX-APUF against MLP}

Mursi \textit{et al.}~\cite{mursi2020fast} demonstrated that MLP is more effective than LR attack, because LR can be understood as a network with a single layer and a single neuron. The purpose of this experiment is to examine the OAX-APUF modeling resilience against MLP and its comparison with XOR-APUF.

\mypara{Setup} 
For OAX-APUF with $x + y + z = 5,6,7$, the number of CRPs required for training is approximately 2-3 times that reported by Mursi \textit{et al.}~\cite{mursi2020fast}. The number 
of training CRPs we used are $100,000$, $650,000$ and $1,360,000$ respectively. While for OAX-APUF with $x+y+z=8$, the training CRPs need to be up to $20,000,000$, which is about 11 times larger than the originally used~\cite{mursi2020fast}. When $x+y+z=5$, the batch size is set to $1,000$, while $x+y+z \ge 5$, the batch size is $10,000$. The epoch number is $200$ when $x+y+z\leq 6$, $50$ for $x+y+z=7$, and $20$ for $x+y+z=8$. XOR-APUF is included here.

For all combinations of MLP attacks, we constantly used $15,000$  CRPs for model prediction during the testing phase to evaluate the attack accuracy.

\mypara{Analysing Results}
The results of the MLP attack on $l$-XOR-APUFs and ($x,y,z$)-OAX-APUFs are summarized in table \ref{MLP-Attack}. Due to the limitation of computational resources, the experiment only attacks OAX-APUF with $x + y + z \leq 8$ and XOR-APUF with $l\leq 8$. The best prediction accuracy of OAX-APUF and XOR-APUF reaches more than 98\%. Compared with LR, MLP requires less CRPs and shorter training time. In addition, the prediction accuracy of MLP is higher than that of LR. One conjecture is that the MLP implemented by \texttt{Keras} does not need to set the hypothesis function according to physical model, which is faster in the modeling attack on OAX-APUF. Concurrently, the MLP architecture uses \textsf{tanh} as the activation function of the hidden layers, which can accurately simulate nonlinear transformations such as OR, AND and XOR. This method thus 
increases the prediction accuracy of the MLP attack on OAX-APUF. According to these results, OAX-APUF using three logical operations cannot perform better than XOR-APUF in resisting MLP attacks.

Wisiol \textit{et al.}~\cite{wisiol2021neural} recently implied that the neural-network based modeling attack can outperform traditional modeling attacks, in particular, the LR that was regarded as the most efficient one against XOR-APUF~\cite{nguyen2019interpose}. This result is also validated in the experiments  of this study using MLP, which is also a typical neural-network.
\begin{table}[]
\caption{MLP attack on XOR-APUF(($x=0,y=0,z=l$)-OAX-APUF) and ($x,y,z$)-OAX-APUF.}
\centering
\label{MLP-Attack}
\resizebox{0.5\textwidth}{!}{
\renewcommand{\arraystretch}{1.2}
\begin{tabular}{|c|c|c|c|c|c|c|c|}
\hline
\textbf{\begin{tabular}[c]{@{}c@{}}Num.\\ APUFs\end{tabular}} & \textbf{\begin{tabular}[c]{@{}c@{}}Training\\ CRPs\end{tabular}} & \textbf{\begin{tabular}[c]{@{}c@{}}Test \\ CRPs\end{tabular}} & \textbf{Epochs}      & \textbf{\begin{tabular}[c]{@{}c@{}}Batch\\ size\end{tabular}} & \textbf{x,y,z} & \textbf{\begin{tabular}[c]{@{}c@{}}Best \\ Pred.Acc\end{tabular}} & \textbf{\begin{tabular}[c]{@{}c@{}}Training \\ Time\end{tabular}} \\ \hline
\multirow{4}{*}{5}                                            & \multirow{4}{*}{100,000}                                         & \multirow{4}{*}{15,000}                                       & \multirow{4}{*}{200} & \multirow{4}{*}{1,000}                                        & 1,2,2          & 98.37\%                                                           & 0.83min                                                           \\ \cline{6-8} 
                                                              &                                                                  &                                                               &                      &                                                               & 2,1,2          & 98.66\%                                                           & 0.62min                                                           \\ \cline{6-8} 
                                                              &                                                                  &                                                               &                      &                                                               & 2,2,1          & 98.92\%                                                           & 0.67min                                                           \\ \cline{6-8} 
                                                              &                                                                  &                                                               &                      &                                                               & 0,0,5          & 98.53\%                                                           & 0.42min                                                           \\ \hline
\multirow{5}{*}{6}                                            & \multirow{5}{*}{650,000}                                         & \multirow{5}{*}{15,000}                                       & \multirow{5}{*}{200} & \multirow{5}{*}{10,000}                                       & 2,2,2          & 99.22\%                                                           & 6.45min                                                           \\ \cline{6-8} 
                                                              &                                                                  &                                                               &                      &                                                               & 1,3,2          & 99.39\%                                                           & 7.2min                                                            \\ \cline{6-8} 
                                                              &                                                                  &                                                               &                      &                                                               & 3,1,2          & 99.37\%                                                           & 6.6min                                                            \\ \cline{6-8} 
                                                              &                                                                  &                                                               &                      &                                                               & 2,3,1          & 99.37\%                                                           & 6.6min                                                            \\ \cline{6-8} 
                                                              &                                                                  &                                                               &                      &                                                               & 0,0,6          & 99.45\%                                                           & 3.1min                                                            \\ \hline
\multirow{5}{*}{7}                                            & \multirow{5}{*}{1,360,000}                                       & \multirow{5}{*}{15,000}                                       & \multirow{5}{*}{50}  & \multirow{5}{*}{10,000}                                       & 3,2,2          & 99.11\%                                                           & 5.69min                                                           \\ \cline{6-8} 
                                                              &                                                                  &                                                               &                      &                                                               & 2,3,2          & 99.25\%                                                           & 6.58min                                                           \\ \cline{6-8} 
                                                              &                                                                  &                                                               &                      &                                                               & 2,2,3          & 99.04\%                                                           & 6.58min                                                           \\ \cline{6-8} 
                                                              &                                                                  &                                                               &                      &                                                               & 2,4,1          & 99.43\%                                                           & 5.58min                                                           \\ \cline{6-8} 
                                                              &                                                                  &                                                               &                      &                                                               & 0,0,7          & 98.94\%                                                           & 4.72min                                                           \\ \hline
\multirow{5}{*}{8}                                            & \multirow{5}{*}{20,000,000}                                      & \multirow{5}{*}{15,000}                                       & \multirow{5}{*}{20}  & \multirow{5}{*}{100,000}                                      & 4,2,2          & 99.57\%                                                           & 40min                                                             \\ \cline{6-8} 
                                                              &                                                                  &                                                               &                      &                                                               & 2,4,2          & 99.59\%                                                           & 38.87min                                                          \\ \cline{6-8} 
                                                              &                                                                  &                                                               &                      &                                                               & 2,2,4          & 99.54\%                                                           & 38.63min                                                          \\ \cline{6-8} 
                                                              &                                                                  &                                                               &                      &                                                               & 2,5,1          & 99.65\%                                                           & 38.83min                                                          \\ \cline{6-8} 
                                                              &                                                                  &                                                               &                      &                                                               & 0,0,8          & 99.42\%                                                           & 40min                                                             \\ \hline
\end{tabular}
}
\end{table}

\begin{figure}[h]
	\centering
	\includegraphics[trim=0 0 0 0,clip,width=0.45\textwidth]{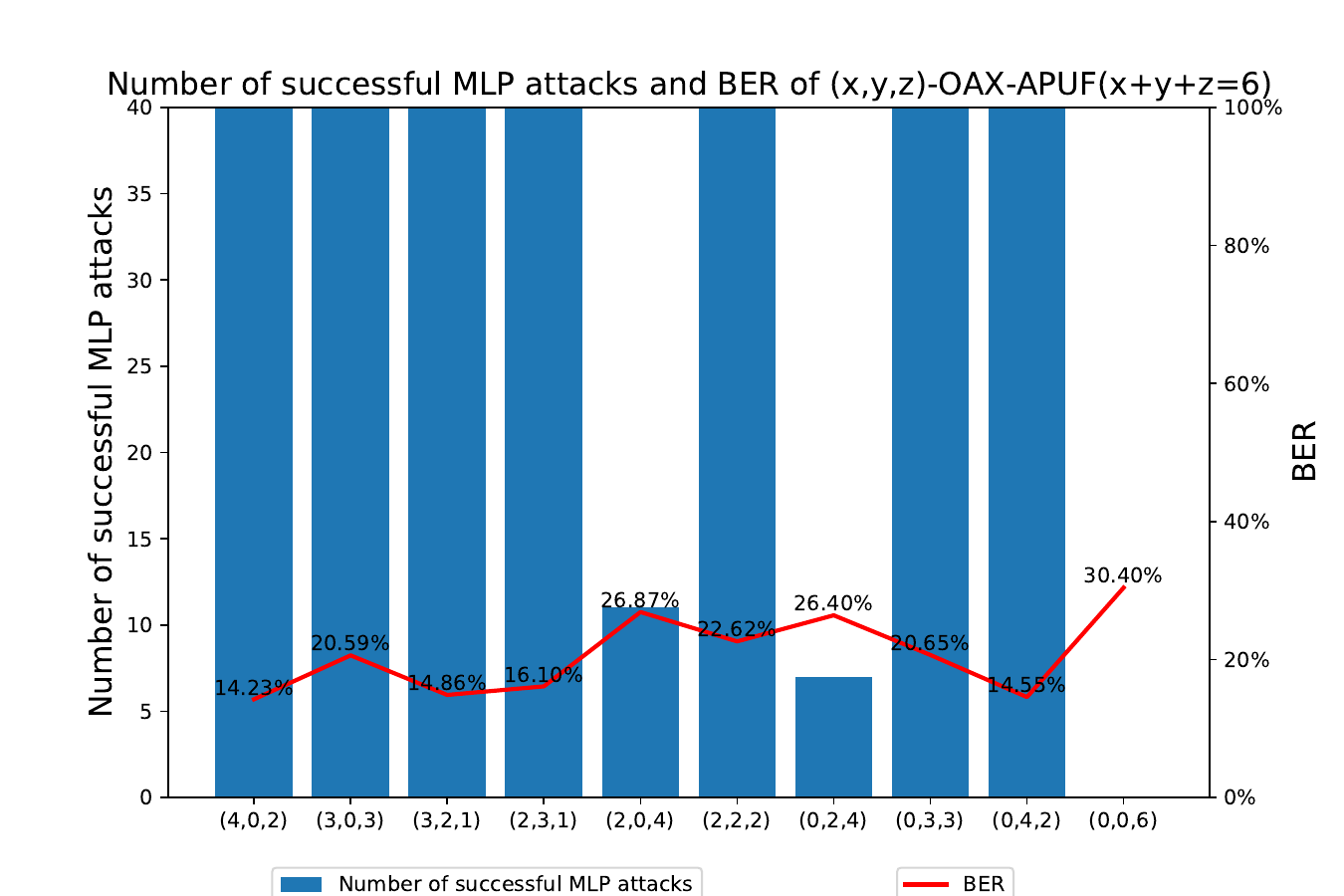}
	\caption{The number of successful MLP attacks and BER of ($x,y,z$)-OAX-APUF when $x+y+z=6$. 
	}
	\label{fig:BER_acc}
\end{figure}

\begin{table}[h]
\centering
\caption{Second hypothesis test results.}
\centering
\label{table:second hypothesis}
\resizebox{0.5\textwidth}{!}{
\renewcommand{\arraystretch}{1.2}
\begin{tabular}{|c|c|c|c|c|c|}
\hline
\textbf{(x,y,z)} & \textbf{\begin{tabular}[c]{@{}c@{}} Training \\ CRPs\end{tabular}} & \textbf{\begin{tabular}[c]{@{}c@{}} Test \\CRPs \end{tabular}} & \textbf{\begin{tabular}[c]{@{}c@{}}Selected  \\ CRPs \end{tabular}} & \textbf{Test Acc} & \textbf{\begin{tabular}[c]{@{}c@{}} Selected CRPs \\ Test Acc \end{tabular}} \\ \hline
(4,0,1)          & 40,000                 & 15,000             & 773                                                                             & 97.31\%           & 84.99\%                    \\ \hline
(4,0,1)          & 100,000                & 15,000             & 773                                                                             & 97.75\%           & 91.85\%                    \\ \hline
(0,4,1)          & 40,000                 & 15,000             & 1,068                                                                           & 96.77\%           & 89.04\%                    \\ \hline
(0,4,1)          & 100,000                & 15,000             & 1,068                                                                           & 97.58\%           & 92.70\%                    \\ \hline
(4,0,2)          & 40,000                 & 15,000             & 942                                                                             & 93.69\%           & 72.19\%                    \\ \hline
(4,0,2)          & 200,000                & 15,000             & 942                                                                             & 95.68\%           & 88.32\%                    \\ \hline
(4,0,2)          & 400,000                & 15,000             & 942                                                                             & 96.37\%           & 88.00\%                    \\ \hline
(4,0,2)          & 650,000                & 15,000             & 942                                                                             & 95.96\%           & 89.60\%                    \\ \hline
(0,4,2)          & 40,000                 & 15,000             & 574                                                                             & 93.90\%           & 62.89\%                    \\ \hline
(0,4,2)          & 200,000                & 15,000             & 574                                                                             & 96.42\%           & 87.80\%                    \\ \hline
(0,4,2)          & 400,000                & 15,000             & 574                                                                             & 96.17\%           & 89.90\%                    \\ \hline
(0,4,2)          & 650,000                & 15,000             & 574                                                                             & 96.49\%           & 90.07\%                    \\ \hline
(0,0,6)          & 400,000                & 15,000             & -                                                                               & 91.47\%           & -                          \\ \hline
(0,0,6)          & 650,000                & 15,000             & -                                                                               & 91.33\%           & -                          \\ \hline
\end{tabular}
}
\end{table}

\textbf{Two Hypotheses.} On the one hand, it is noticed that if the $x$ or $y$ is large (i.e., 4), the $r_{\rm or}$ or $r_{\rm and}$ will be highly biased (as shown in Table~\ref{tab:uniformity}), which indicates that the MLP may not need to learn the sub-block of $x$-OR-APUF or $y$-AND-APUF within the ($x,y,z$)-OAX-APUF. $r_{\rm or}$ to be `1' or $r_{\rm and}$ can thus equal to `0'. Therefore, the modeling resilience of ($x,y,z$)-OAX-APUF will be lower when $x$/$y$ increases. On the other hand, as the MLP could blindly assign `1'/`0' to $x$-OR-APUF/$y$-AND-APUF sub-block, it could not essentially learn the $x$-OR-APUF/$y$-AND-APUF and exhibit higher error rate when $r_{\rm or}$ is essentially `0' or $r_{\rm and}$ is essentially `1'.

\textbf{First Hypothesis Test.}
To test the former hypothesis, we set up the following experiments.

We choose $x+y+z=6$ as an example, and intentionally use fewer CRPs, $40,000$ CRPs, for MLP attacks on each of these combinations. We repeat the attacks per combination 40 times and count the number of successful attacks. When the attack accuracy is higher than the reliability (1-BER), the attack is regarded as successful. 

As shown in Fig.~\ref{fig:BER_acc}, the combination with $z$ less than 4 can be broken by MLP 40 times. When $z = 4$, the number of MLP successful attacks is less, but $40,000$ CRPs cannot break 6-XOR-APUF. These results prove the first hypothesis. In this context, $z$ should be increased to increase the modeling resilience to MLP attacks. As one of main impetus of OAX-APUF is to defeat CMA-ES enabled reliability attacks, it is preferred to retain $x$ or $y$ to be the smaller number of two and use the remaining APUFs contributing to the $z$-XOR-APUF.

\textbf{Second Hypothesis Test.} We choose \{($4,0,1$), ($0,4,1$), ($4,0,2$), ($0,4,2$)\} combined OAX-APUF for evaluation. For $4$-OR-APUF (first and third combinations), we select the challenges when their responses are all `0's for all four APUFs in this $4$-OR-APUF block. For $4$-AND-APUF (second and fourth combinations), we select the challenges when their responses are all `1's for all four APUFs in this $4$-OR-APUF block.

The results are shown in Table~\ref{table:second hypothesis}. Even though the response accuracy predicted given random unseen challenges is high when using $40,000$ training CRPs to learn the OAX-APUF, the MLP cannot predict the selected CRPs accurately. For example, the attack merely exhibits 72\% and 62\% accuracy for selected challenges of the ($4,0,2$)-OAX-APUF and ($0,4,2$)-OAX-APUF, respectively, which validates the second hypothesis that the MLP is `lazy' to learn the underlying OR-APUF or AND-APUF sub-blocks when the $x$/$y$ is increased, where it simply deems the response $r_{\rm or}$/$r_{\rm and}$ to be `1'/`0'. To accurately learn the underlying OR-APUF or AND-APUF sub-blocks, the MLP does require more CRPs for training. More specifically, to achieve an accuracy of close to 90\% for the selected challenges, the required number of CRPs for training  the ($4,0,2$)-OAX-APUF and ($0,4,2$)-OAX-APUF approaches that of training the XOR-APUF (($0,0,6$)-OAX-APUF).

Assuming that the attacker only uses a small CRP set (i.e.,$40,000$) to train the ($4,0,2$)-OAX-APUF and ($0,4,2$)-OAX-APUF, the server can strategically use the selected challenges for authentication. This process is possible if we assume that the server can access each APUF and construct an accurate model for each and then disable access, leaving the attacker who can only access the OAX response. Utilization of this asymmetric access ability to perform authentication has been leveraged in~\cite{yu2016lockdown,9091218}. In this context, the attacker is forced to use more CRPs for training the ($0,4,2$)-OAX-APUF to a large extent, which again is comparable the number to break the $l$-XOR-APUF with $x+y+z = l$.

Nonetheless, as an intuitive means of increasing the OAX-APUF's modeling resilience while making it exhibit great resilience to other ML techniques based attacks (especially the reliability based CMA-ES attack), using least number of APUFs (i.e., 2) for OR or AND logic operations appears to be preferred.

\section{Related Works}\label{sec:relatedwork}

In addition to the well-studied XOR-APUF, there have been multiple works about APUF recomposition that have been reported recently~\cite{yu2016lockdown,sahoo2017multiplexer,nguyen2019interpose,zhang2020set,9091218,gu2020modeling,avvaru2020homogeneous}. Among these works, only ($x,y$)-iPUF~\cite{nguyen2019interpose}, MPUF~\cite{sahoo2017multiplexer}, and XOR-FF-APUF~\cite{avvaru2020homogeneous} require neither additional security primitives nor memory elements and thus are relevant to the proposed OAX-APUF, which all still have an open challenge-response interface. Protected interfaces are discussed in Section~\ref{sec:protectInterface}. 

\mypara{\bf ($x,y$)-iPUF} 
In terms of modeling resilience against LR attack, ($x,y$)-iPUF is similar to the ($x/2+y$)-XOR-APUF~\cite{nguyen2019interpose}. This modeling complexity is reduced to $\text{max}({x,y}$)-XOR-APUF by a recent study~\cite{wisiol2020splitting} that models the upper $x$ APUF components first and then the lower $y$ APUF components in $(x,y)$-iPUF in a divide-and-conquer manner---this attack comes with a cost for more CRPs and computational resources~\cite{wisiol2020splitting}. For the reliability-based attack, the CMA-ES cannot build models for $x$ upper APUFs in $(x,y)$-iPUF due to their unreliability contribution to $(x,y)$-iPUF being smaller than those $y$ lower APUFs.

Hybrid LR-reliability attack can successfully break $(x,y)$-iPUF, such as $(1,10)$-iPUF. Because the weights of $y$ lower APUFs is ambiguous, inverting one half of a $y$ lower APUF weight vector does not change its reliability loss~\cite{tobisch2021combining}. Therefore, Eq.~\ref{eq:Hybrid LR-reliability loss} needs to be modified. Tobisch \textit{et al.}~\cite{tobisch2021combining} proposed a differentiable model of $(x,y)$-iPUF, which is exploited in the hybrid LR-reliability attack on $(x,y)$-iPUF. They used this single-pass approach to successfully learn up to $(1,10)$ and $(4,4)$-iPUFs. Hybrid LR-reliability attack can also be used in multi-pass attack~\cite{wisiol2020splitting} to attack $(x,y)$-iPUF (e.g., $(4,4),(5,5),(6,6),\text{and}~(7,7)$-iPUFs). 

\mypara{\bf MPUF} MPUF is also a type of APUF recomposition~\cite{sahoo2017multiplexer} that has three variants: ($n,k$)-MPUF, ($n,k$)-rMPUF and ($n,k$)-cMPUF, where $n$ stands for the number of stages of the underlying APUF. Taking the basic ($n,k$)-MPUF as an example, a $2^k$-to-1 multiplexer (MUX) with a selection vector length $k$ is used where $k$ responses from $k$ APUFs are used as $k$ selection inputs of the MUX. Responses of $2^k$ APUFs are the data inputs of the MUX, so that out of $2^k$ APUFs responses will be selected as the response output of the ($n,k$)-MPUF. All $2^k + k$ APUFs share the same challenge with a challenge length of $n$. The ($n,k$)-rMPUF is an improvement over ($n,k$)-MPUF to achieve better robustness against reliability-based ML attacks, which requires $2^{k+1} - 1$ in total. ($n,k$)-cMPUF is another variant over ($n,k$)-MPUF to resist linear cryptanalysis, which requires $2^{k-1} + k$ in total.

Compared to IPUF, XOR-APUF and OAX-APUF, the MPUF recomposition requires not only significantly more APUFs but also a lot of MUXs (e.g., $2^k$-to-1 MUXs are first decomposed into many $2$-to-1 MUXs for implementation). In~\cite{sahoo2017multiplexer}, ($n,3$)-rMPUF with $n=64$ achieves better reliability and modeling resilience than $10$-XOR-APUF. In~\cite{shi2019approximation,alamro2021machine}, the security of ($n,3$)-rMPUF ($n=64$) has been broken. It has been also shown that the ($64,6$)-rMPUF and ($32,7$)-rMPUF are breakable~\cite{alamro2021machine}. The implementation of a further scaled ($64,7$)-rMPUF and ($32,8$)-rMPUF requires at least 255 64-stage APUFs and 511 32-stage APUFs, respectively, and the high APUF number requirement prevents its practicality as  a strong PUF candidate under an open access interface in practice.

\mypara{\bf XOR-FF-APUF} 
FF-APUF has an additional intermediate arbiter that produces an intermediate response. This intermediate response is used as a challenge bit for one following stage, which can also be more, in the APUF. Multiple intermediate responses can be provided with multiple intermediate arbiters. In~\cite{avvaru2020homogeneous}, Avvaru \textit{et al.} investigate the security and reliability of the $x$-XOR-FF-APUF, where the responses from $x$ FF-APUFs are XORed. When $x$ is fixed, e.g., $x=5$, $x$-XOR-FF-APUF  achieves better security against modeling attack than the $l$-XOR-APUF. When the position of the intermediate arbiter and its response insertion of $x$-FF-APUFs are the same, $x$-XOR-FF-APUF is termed as homogeneous $x$-XOR-FF-APUF. In other cases, $x$-XOR-FF-APUF is called heterogeneous $x$-XOR-FF-APUF. The security of the heterogeneous $x$-XOR-FF-APUF is shown to be stronger than the homogeneous one.

\section{Discussion}\label{sec:discussion}
\subsection{OAX-PUF Evaluation Summary}

For \textbf{RQ 1}, we have formulated the uniformity and reliability of the ($x,y,z$)-OAX-PUF, which are generic to any underlying PUF types used to recomposite the OAX-PUF. The uniformity is not deteriorated, and remains at 50\%. The reliability has been improved compared to the pure XOR based recomposition (detailed in Table~\ref{tab:CMA-ES}). Notably, each PUF's unreliability contribution to the recomposited OAX-PUF has been disrupted, which is the fundamental reason to prevent the powerful reliability based CMA-ES attack.

For \textbf{RQ 2}, we take the APUF as  the underlying PUF to recompose the OAX-PUF as a case study and compare its modeling resilience with XOR-APUF. Based on empirical results against four powerful modeling attacks, the OAX-APUF can still be effectively modeled except by defeating the CMA-ES attack. Inadvertently, OAX-APUF hardens the modeling attack, to some extent, in terms of attacking computation time when similar attacking accuracy is achieved compared to XOR-APUF against hybrid LR-reliability and LR attack.

\subsection{Protected Challenge-Respponse Interface}\label{sec:protectInterface}
Clearly, the security of a PUF system can be ensured if additional security blocks are used to protect the challenge-response interface~\cite{gao2016obfuscated,yu2016lockdown,herder2016trapdoor,gao2017puf,9091218}, such as the Lockdown-PUF with additional RNG, and TREVERSE with additional Hash~\cite{gao2016obfuscated,yu2016lockdown,herder2016trapdoor,gao2017puf,9091218}. In this context, leveraging either of the four recomposited PUFs as detailed in Section~\ref{sec:relatedwork}, including the presented OAX-APUF, can be complementary via proper adoption. For example, OAX-APUF can replace XOR-APUF that is used in Lockdown-PUF system~\cite{yu2016lockdown} to ensure a much higher number of secure authentication rounds. The reason lies on the fact that the OAX-APUF has higher reliability while demonstrate comparable modeling resilience than XOR-APUF (note that the asymmetric access strategy can be leveraged to improve modeling resilience to MLP attacks as in the second hypothesis test in Section~\ref{sec:evaluateMLP}). Thus, each authentication round requires fewer CRPs to be exposed. Specifically, whenever the XOR-APUF used in the Lockdown-PUF system has exposed a pre-estimated number of usable CRPs, these CRPs can be used to build an accurate model of the XOR-APUF. In such a case, the lockdown-PUF instance can no longer be used for the upcoming CRP-based authentication and thus must be discarded. Either increasing the pre-estimated number of usable CRPs or reducing the exposed CRPs in a single authentication round can improve the total number of authentication rounds.

\subsection{Types of PUF Recomposition} As highlighted in~\cite{gao2020physical,ruhrmair2013puf}, the key insight behind the PUF recomposition is that one type of modeling attack typically tends to be efficient only on one type of PUF topology. By combining different types of PUF typologies together, it is expected to improve the recomposited PUF's resilience when each of these modeling attacks is individually mounted. In this work, we have followed this insight and demonstrated the enhanced modeling resilience with OAX-APUF. Instead of solely relying on the XOR bitwise operation to add nonlinearity to OAX-APUF, we have also used OR and AND logic operations as differing means of nonlinearity injection. 
The proposed OAX-APUF exhibits better reliability. Although current PUF recomposition is still threatened by ever evolved modeling attacks, it is expected a proper recomposition should markedly harden these attacks (e.g., later discussed in Section~\ref{sec:hetero}). This is esimilar to a single round of substitution, and the permutation network in the block cipher design yields weak security, but properly applying more rounds with careful organization can lead to a secure design. Beyond OR, AND, and XOR enabled logic operations, future work can investigate a combination of other simple arithmetic operations, such as summation and subtraction, as additional means of nonlinearity injection and may combine them. 

\subsection{Homogeneous and Heterogeneous Underlying PUFs}\label{sec:hetero} 
An underlying PUF is a basic APUF and can be replaced by an FF-APUF to enhance the modeling resilience of recomposited PUFs, because a single FF-APUF has better modeling resilience than a single basic APUF~\cite{avvaru2020homogeneous}. When the same type of underlying PUFs are used for recomposition, this can be treated as homogeneous PUF recomposition. One can essentially employ different PUFs as underlying PUFs, which can be referred to as heterogeneous PUF recomposition. For example, in the OAX-PUF, we can use not only the APUF but also the FF-APUF for recomposition, where both APUF and FF-APUF still share the same challenge. It has demonstrated a better modeling resilience of a heterogeneous XOR-FF-APUF compared with homogeneous XOR-FF-APUF in~\cite{avvaru2020homogeneous}. 

Overall, future recomposited PUF variants can consider both different nonlinearity injection operations and heterogeneous underlying PUFs.

\section{Conclusion} \label{sec:conclusion}
In this paper, we explicitly design and evaluate a new lightweight recomposited PUF, ($x,y,z$)-OAX-PUF. Its two key performances (i.e., reliability and uniformity) are formulated and validated. Overall, ($x,y,z$)-OAX-PUF exhibits enhanced reliability compared with $l$-XOR-PUF without sacrificing any uniformity given $l=x+y+z$. Considering the APUF as a case study to examine the OAX-PUF modeling resilience, modeling resilience is not as high as expected, which can still be broken by newly evolved modeling attacks, in particular, MLP attack and hybrid LR-reliability based attack. Conversely, OAX-APUF requires a longer time for most attack types compared to XOR-APUF and successfully defeats the CMA-ES attack. Future research can investigate combinations of varying recomposition types and leveraging heterogeneous underlying PUFs (e.g., FF-APUF) and 
achieve modeling resilience for recompositing PUFs.  

\section{Acknowledgment}
We acknowledge support from the National Natural Science Foundation of China (62002167 and 61901209) and National Natural Science Foundation of JiangSu (BK20200461).


\bibliographystyle{IEEEtran}

\clearpage
\begin{appendices}
      \section{}
       \subsection{Synthesizing APUF with RO Frequencies}
      
\begin{figure}[h]
	\centering
	\includegraphics[trim=0 0 0 0,clip,width=0.45\textwidth]{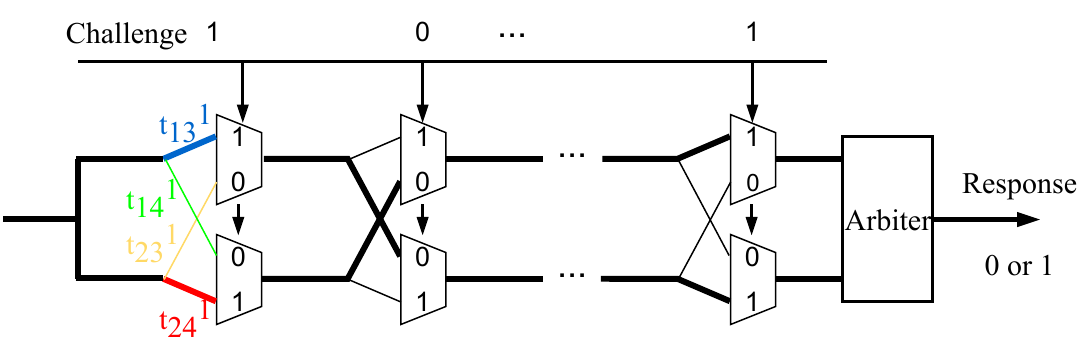}
	\caption{Example of one APUF with two signal transmission paths.
	}
    \label{fig:apuf}
    \end{figure}

      Benefiting from two large-scale silicon ROPUF dataset (they are measured under varying temperature/voltage), we follow~\cite{gao2014highly,9091218} that treat the inverse of the RO frequency as gate level delay, thus can leverage the delay measurements of the ROPUF to stand for delay at each stage of the APUF. The constructed APUF in this manner is termed as RO-APUF.
    More specifically, the RO-APUF is constructed as follows:
      \begin{itemize}
    \item Get the reciprocal of RO frequencies to serve as the path delay of APUF: the reciprocal of four RO frequencies are used as the time delay of the $i$-th stage of APUF: t$_{13}^{i}$,t$_{14}^{i}$,t$_{23}^{i}$,t$_{24}^{i}$. Simulating a 64 stage APUF requires 256 RO frequencies.
    \item As shown in Fig.~\ref{fig:apuf}, if $c[i]=1$, the two signals propagate from 1 to 3 and 2 to 4, respectively. If $c[i]=0$, for the delay of the $i_{\rm th}$ stage, select t$_{14}^{i}$ and t$_{23}^{i}$. 
    \item delay\_cross$^{i}$ and delay\_uncross$^{i}$ are used to represent the time delay of $i_{\rm th}$ stage. $delay\_cross^{i}=t_{14}^{i}-t_{23}^{i}$, $delay\_uncross^{i}=t_{13}^{i}-t_{24}^{i}$.
   \item Form $w$ vector as Eq.\ref{wform}:
    \begin{equation}
    \begin{split}
    \label{wform}
        &w[1]=(delay\_uncross^{1}-delay\_cross^{1})/2, \\
        &w[65]=(delay\_uncross^{64}+delay\_cross^{64})/2,\\
        &w[i]=(delay\_uncross^{i-1}+delay\_cross^{i-1}\\
        &+delay\_uncross^{i}-delay\_cross^{i})/2,
    \end{split}
    \end{equation}
    here $i=2,3,...,63.$
    
    \item Compute the response of a given challenge according to Eq.\ref{con:delta},Eq.\ref{con:feature vector},Eq.\ref{con:responses} .
    % \item $d_{\rm up}$ and $d_{\rm bottom}$ are used to represent the time delay of the top and bottom signals respectively, and the initial value is set to be 0.
    % \item As shown in Fig.~\ref{fig:apuf}, if $c[i]=1$, the two signals propagate from 1 to 3 and 2 to 4, respectively, at the $i_{\rm th}$ stage, then $d_{up}=d_{up}+t_{13}^{i}$,~$d_{bottom}=d_{bottom}+t_{24}^{i}$. If $c[i]=0$, for the delay of the $i$-th stage, select t$_{14}^{i}$ and t$_{23}^{i}$, that is, $d_{up}=d_{up}+t_{14}^{i}$,~$d_{down}=d_{down}+t_{23}^{i}$. The delay difference of the top and bottom path delay $\Delta=d_{up}-d_{bottom}$, if $\Delta \textgreater 0$, $r=0$, otherwise, $r=1$.
   \end{itemize}
    
   \begin{figure}[h]
    	\centering
    	\includegraphics[trim=0 0 0 0,clip,width=0.45\textwidth]{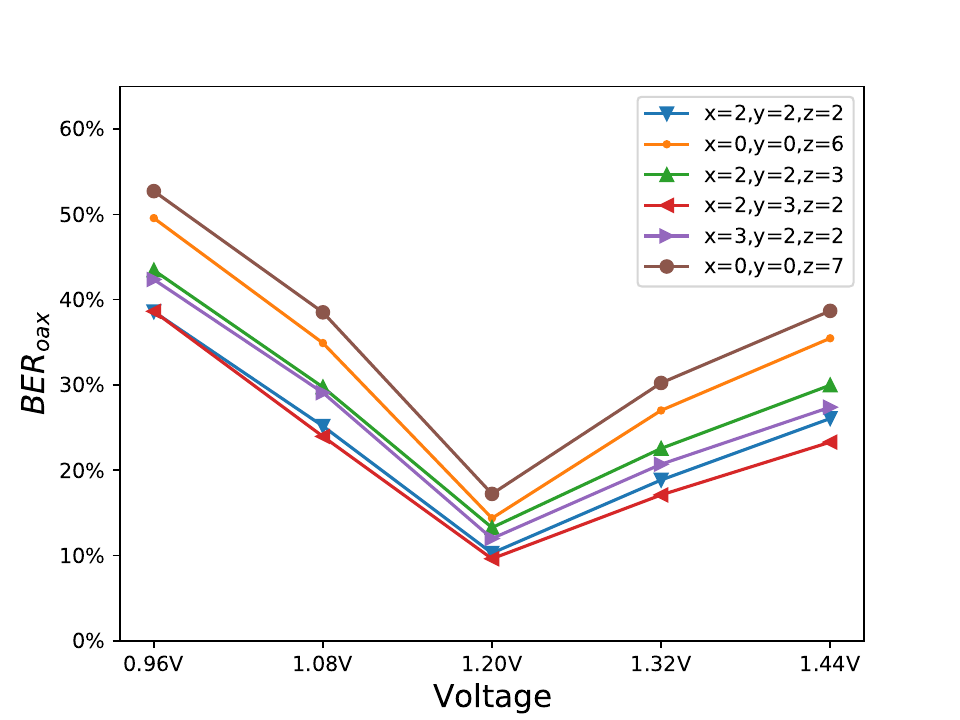}
    	\caption{BER of ($x,y,z$)-OAX-APUF$_{RO}$ ($x+y+z=6,x+y+z=7$) at different voltages (Virginia Tech dataset).
    	}
      \label{fig:ber_vol2010}
    \end{figure}

    \begin{figure}[h]
    	\centering
    	\includegraphics[trim=0 0 0 0,clip,width=0.45\textwidth]{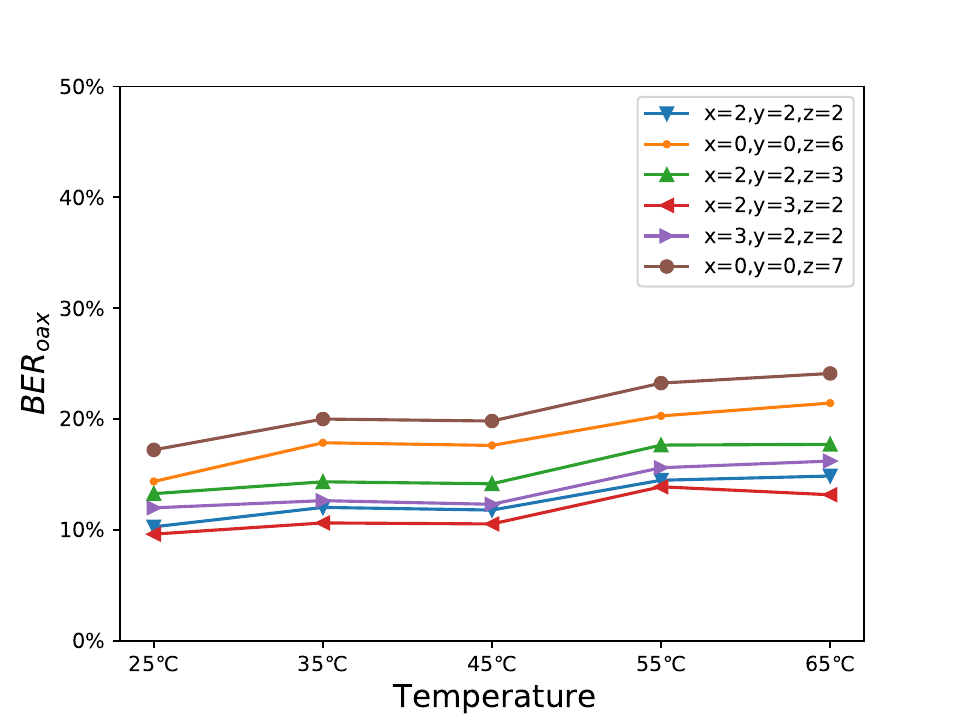}
        \caption{BER of ($x,y,z$)-OAX-APUF$_{RO}$ ($x+y+z=6,x+y+z=7$) at different temperatures (Virginia Tech dataset).
        }
      \label{fig:ber_tem2010}
    \end{figure}

    \begin{figure}[h]
    	\centering
    	\includegraphics[trim=0 0 0 0,clip,width=0.45\textwidth]{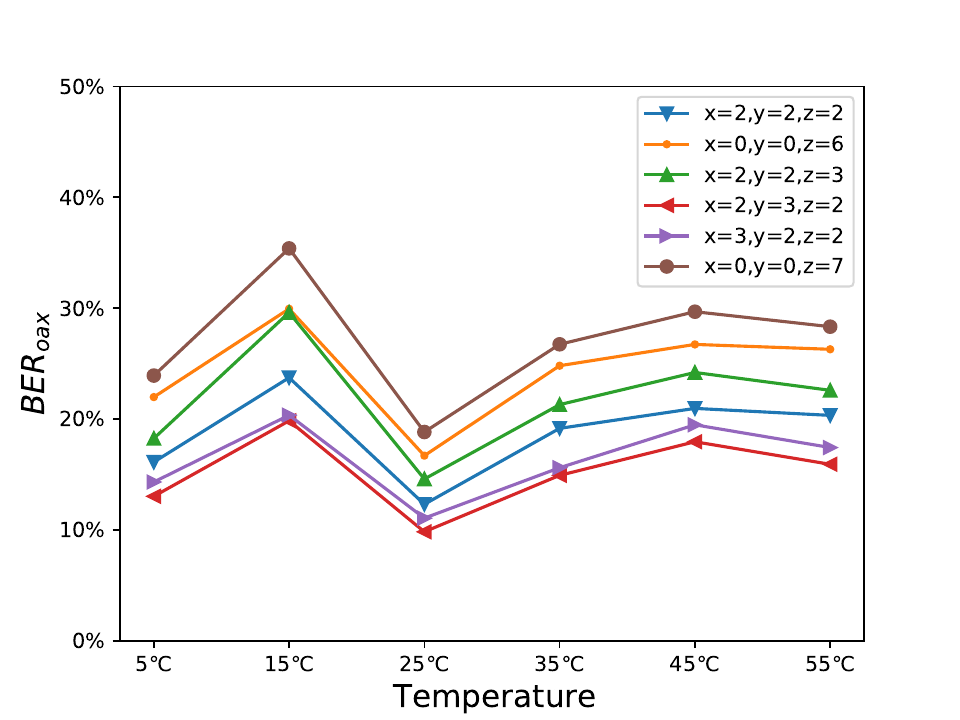}
    	\caption{BER of ($x,y,z$)-OAX-APUF$_{RO}$ ($x+y+z=6,x+y+z=7$) at different temperatures (HOST2018 dataset).
    	}
      \label{fig:ber_tem2018}
    \end{figure}

    \begin{figure}[h]
    	\centering
    	\includegraphics[trim=0 0 0 0,clip,width=0.45\textwidth]{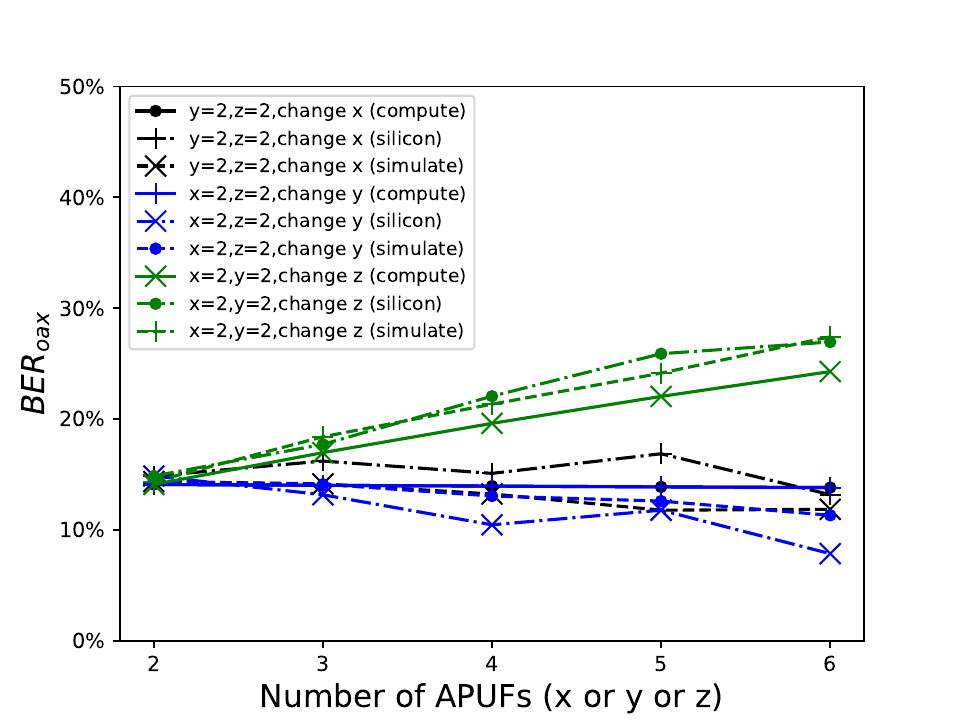}
    	\caption{BER of ($x,y,z$)-OAX-APUF. For the silicon measurement based on RO synthesized APUF, the regenerated response is at (65\textcelsius, 1.20V)---this gives the worst BER when temperatures changed---and the e1nrolled response done at (25\textcelsius, 1.20V) (Virginia Tech Dataset).
    	}
      \label{fig:ber_ten65}
    \end{figure}

    \begin{figure}[!h]
    	\centering
    	\includegraphics[trim=0 0 0 0,clip,width=0.45\textwidth]{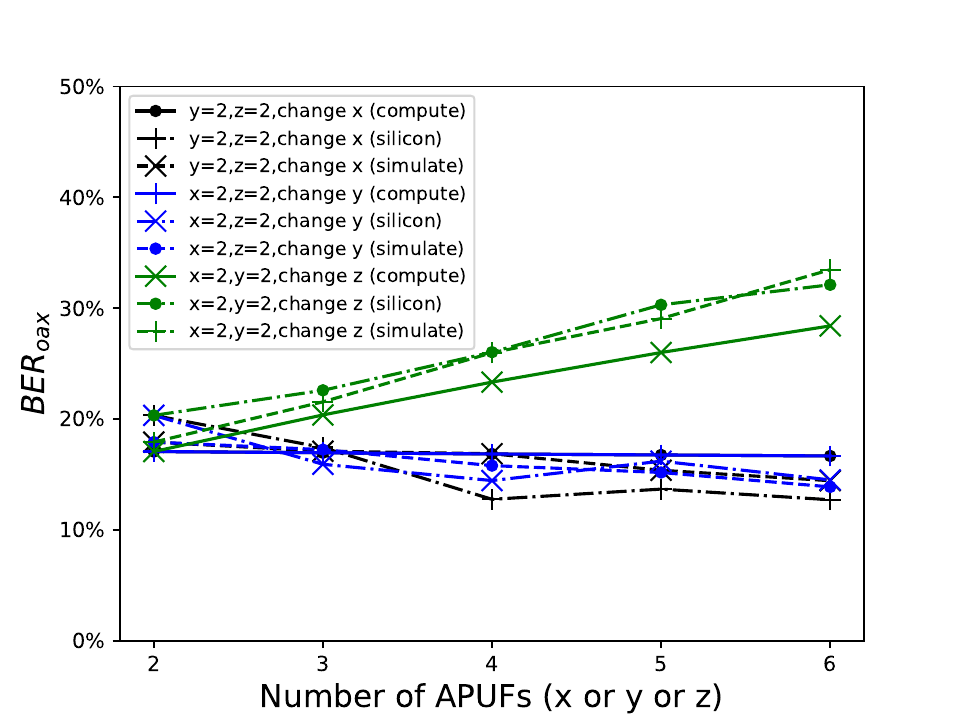}
    	\caption{BER of ($x,y,z$)-OAX-APUF.
	For the silicon measurement based on RO synthesized APUF, the regenerated response is at 55\textcelsius ~and the enrolled response done at 25\textcelsius~ (HOST2018 Dataset).
    	}
      \label{fig:ber_ten55}
    \end{figure}
      \subsection{Public Datasets}
       We use two public  ROPUF datasets to construct RO-APUF: Virginia Tech~\cite{maiti2010large} and HOST2018~\cite{hesselbarth2018large}. As for the Virginia ROPUF dataset, five ROPUFs are implemented across five Spartan3E S500 FPGA boards. Each FPGA implements one ROPUF that consists of 512 ROs~\cite{9091218}. Each RO frequency measurement is repeated 100 times under the operating voltages of 0.96V, 1.08V, 1.20V, 1.32V, and 1.44V at a fixed temperature of 25\textcelsius~to capture supply voltage influences~\cite{9091218}. Similarly, each RO frequency is also evaluated 100 times under 35\textcelsius, 45\textcelsius, 55\textcelsius, and 65\textcelsius, with a fixed supply voltage of 1.20V, to reflect influence from temperature changes~\cite{9091218}. HOST2018 provides raw data of 217 Xilinx Artix-7 XC7A35T FPGAs, each containing a total of 6592 ROs, comprised of six different routing paths with 550 to 1696 instances per type~\cite{hesselbarth2018large}. Each RO frequency is evaluated 100 times under 5\textcelsius, 15\textcelsius, 25\textcelsius, 35\textcelsius, 45\textcelsius, and 55\textcelsius.
       
    \subsection{The BER results of OAX-APUF$_{RO}$}
          For Virginia Tech, we take the response of OAX-APUF$_{RO}$ at (25\textcelsius, 1.20V) as reference, and repeatedly measure the responses at other temperatures/voltages for 11 times, so as to evaluate the BER of OAX-APUF$_{RO}$ under different environment. As shown in Fig.~\ref{fig:ber_vol2010}, when the voltage deviates from 1.20V, the BER of OAX-APUF$_{RO}$ increases, and the worst BER occurs at 0.96V. From Fig.~\ref{fig:ber_tem2010}, we can see that when the temperature deviates from the nominal temperature of 25 \textcelsius, the BER of OAX-APUF$_{RO}$ also increases.

      For HOST2018, we take the response of OAX-APUF$_{RO}$ at 25\textcelsius~as a reference, and repeatedly measure the responses at other temperatures/voltages for 11 times, so as to evaluate the BER of OAX-APUF$_{RO}$. Fig.\ref{fig:ber_tem2018} shows the BER of several OAX-APUF$_{RO}$s simulated by the HOST2018 dataset at different temperatures. When the temperature is less than 25\textcelsius, the BER of OAX-APUF$_{RO}$ is the highest at 15\textcelsius. When the temperature is greater than 25\textcelsius~the BER gradually increases.

      To validate the formulated BER, we have now included the OAX-APUF$_{RO}$ in additiont to the simulated OAX-APUF. For Virginia Tech, we choose the worst-case BER at (25\textcelsius, 0.96V) and (65\textcelsius, 1.20V)---the former has been detailed in Section 3.4, so here we show results of later. When the operating condition of regenerating response is (65\textcelsius, 1.20V), the average BER of APUF$_{RO}$s is about 0.04. So 0.04 is used for calculation as well as simulated APUF to be consistent for apple-to-apple comparison, which BER results of OAX-APUF$_{RO}$s are detailed in Fig.~\ref{fig:ber_ten65}. As for the HOST2018, the average BER of APUF$_{RO}$ is 0.05 when response is regenerated at 55\textcelsius. So that 0.05 BER is used for equation calculation and simulated APUFs, which BER results of OAX-APUF$_{RO}$s are detailed in  Fig.~\ref{fig:ber_ten55}. From both Fig.~\ref{fig:ber_ten65} and Fig.~\ref{fig:ber_ten55}, we can see that BER of both silicon and simulated OAX-APUF$_{RO}$s match our formulated BER well.
  \end{appendices}

\end{document}